\documentclass[pre,twocolumn,superscriptaddress,showpacs,pdfoutput=1]{revtex4-1}
\usepackage{graphicx}% Include figure files
\usepackage{epsfig}% Include figure files
\usepackage{amssymb,amsmath,amsfonts,hyperref}
\usepackage{wasysym}
\usepackage{bm}
\usepackage{latexsym}
\usepackage{eucal}
\usepackage{float}
\usepackage{verbatim}
\usepackage[normalem]{ulem}
\usepackage{epstopdf}
\usepackage{color}
\usepackage{braket}
\usepackage{mathrsfs}

\newcommand{\be}{\begin{equation}}
\newcommand{\ee}{\end{equation}}

\newcommand{\eea}{\end{eqnarray}}

\begin{document}

\title{Unconventional U(1) to $\mathbf{Z_q}$ cross-over in quantum and classical ${\bf q}$-state clock models}

\author{Pranay Patil}
\email{pranayp@bu.edu}
\affiliation{Department of Physics, Boston University, 590 Commonwealth Avenue, Boston, Massachusetts 02215, USA}
\affiliation{Laboratoire de Physique Th\'eorique, Universit\'e de Toulouse, CNRS, UPS, France}

\author{Hui Shao}
\email{huishao@bnu.edu.cn}
\affiliation{Center for Advanced Quantum Studies, Department of Physics,
Beijing Normal University, Beijing 100875, China}

\author{Anders W. Sandvik}
\email{sandvik@bu.edu}
\affiliation{Department of Physics, Boston University, 590 Commonwealth Avenue, Boston, Massachusetts 02215, USA}
\affiliation{Beijing National Laboratory for Condensed Matter Physics and Institute of Physics, Chinese Academy of Sciences, Beijing 100190, China}

\begin{abstract}
We consider two-dimensional $q$-state quantum clock models with quantum fluctuations connecting states with all-to-all clock transitions with different
choices for the matrix elements. We study the quantum phase transitions in these models using quantum Monte Carlo simulations and finite-size scaling,
with the aim of characterizing the cross-over from emergent U(1) symmetry at the transition (for $q \ge 4$) to $Z_q$ symmetry of the ordered state.
We also study classical three-dimensional clock models with spatial anisotropy corresponding to the space-time anisotropy of the quantum systems.
The U(1) to ${Z_q}$ symmetry cross-over in all these systems is governed by a so-called dangerously irrelevant operator. We specifically study
$q=5$ and $q=6$ models with different forms of the quantum fluctuations and different anisotropies in the classical models. In all cases, we find
the expected classical XY critical exponents and scaling dimensions $y_q$ of the clock fields. However, the initial weak violation of
the U(1) symmetry in the ordered phase, characterized by a $Z_q$ symmetric order parameter $\phi_q$, scales in an unexpected
way. As a function of the system size (length) $L$, close to the critical temperature $\phi_q \propto L^p$, where the known value of the exponent
is $p=2$ in the classical isotropic clock model. In contrast, for strongly anisotropic classical models and the quantum models we find $p=3$.
For weakly anisotropic classical models we observe a cross-over from $p=2$ to $p=3$ scaling. The exponent $p$ directly impacts the exponent $\nu'$
governing the divergence of the U(1) to $Z_q$ cross-over length scale $\xi'$ in the thermodynamic limit, according to the relationship $\nu'=\nu(1+|y_q|/p)$,
where $\nu$ is the conventional correlation length exponent. We present a phenomenological argument for $p=3$ based on an anomalous renormalization
of the clock field in the presence of anisotropy, possibly as a consequence of topological (vortex) line defects. Thus, our study points to an
intriguing interplay between conventional and dangerously irrelevant perturbations, which may affect also other quantum systems with emergent
symmetries.
\end{abstract}

\maketitle

\section{Introduction}

Emergent symmetries in quantum critical systems have been the subject of numerous recent discussions in condensed matter physics
\cite{EM1,EM2,EM3,EM4,Senthil,Sandvik07,Lou09,Pujari13,Nahum,NahumX2,NDQC,NahumX,Liu19,CDmr,Sato17,Zhao19,Serna,Takahashi20,Zhang18}
and more generally in quantum field theory \cite{Gorbenko18,Gazit18,Yu19,Szabo20} and the conformal bootstrap \cite{Nakayama16,Chester20,He20}.
An emergent symmetry is often associated with a length scale $\xi'$ above which the symmetry is violated as a consequence of a so-called dangerously
irrelevant (DI) perturbation close to a continuous phase transition \cite{Fisher,Nelson,Amit}. This cross-over scale diverges faster upon
approaching the critical point than the conventional correlation length $\xi$. The scenario of deconfined quantum-critical points in two-dimensional
(2D) quantum magnets is a prominent example \cite{Senthil,Sandvik07,Lou09,Pujari13,Nahum,NahumX2,NDQC,NahumX,Liu19,CDmr}, where a DI operator leads to
emergent U(1) symmetry \cite{Levin} of the $Z_4$ \cite{Sandvik07,Lou09,Jiang08} or $Z_3$ \cite{Pujari13,Pujari15} order parameter of a dimerized
(valence-bond solid, VBS) phase. Here the lattice itself imposes the discreteness of the microscopic order parameter (translating to tripled
or quadrupled monopoles in the field theory \cite{Murthy}), but macroscopically
the order appears U(1)-like when coarse grained on length scales below $\xi'$.

\begin{figure}[b]
\vskip-2mm
\includegraphics[width=0.75\hsize]{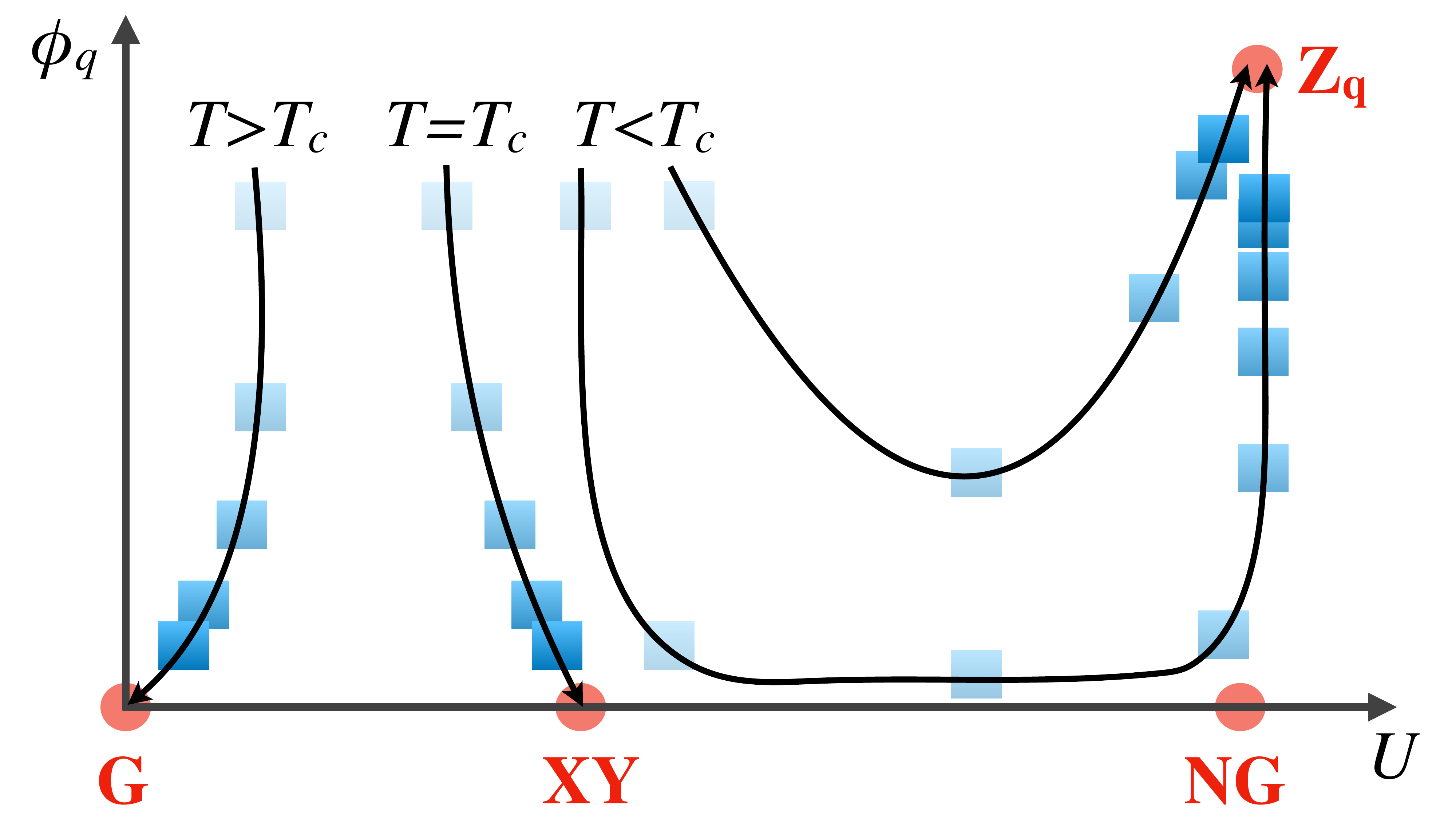}
\caption{Schematic RG flow diagram for the 3D clock model \cite{Huinew} in the space of the Binder cumulant $U$
and the $Z_q$ order parameter $\phi_q$. Darker squares denote larger systems and the flow is from small to large sizes. The fixed points
are the paramagnetic Gaussian point G (reached for $T>T_c$), the critical XY point (reached at $T_c$), the U(1) symmetry breaking NG point
(which the system approaches closely if $T<Tc$ in the neighborhood of $T_c$), and the ordered $Z_q$ symmetry breaking point (reached asymptotically
for all $T<T_c$).}
\label{fflows}
\end{figure}

DI perturbations of this kind are better known from classical models, with a prototypic example being three-dimensional (3D) $q$-state clock models with
$q \ge 4$ \cite{DIO1,Ueno,Chubukov,Miyashita,Oshikawa,Hove,LouSan,Hasenbusch11,Okubo,Zhitomirsky,LenDel,Banerjee,Hasenbusch19,Huinew}. As illustrated in
Fig.~\ref{fflows}, the discreteness of the allowed spin angles, or the presence of a soft $q$-fold symmetric potential, constitutes an irrelevant perturbation
of the XY model at the critical point, where the angular fluctuations of the coarse-grained order parameter become more uniform with increasing length scales.
The ``dangerous'' aspect of the problem pertains to the ordered phase, where the clock perturbation becomes relevant and the symmetry crosses over from U(1)
to $Z_q$ at the second length scale $\xi'$ (and on a more technical level to a non-analytic scaling function of the DI field in the
original context were the concept was developed \cite{Fisher,Nelson,Amit}).

In this paper we introduce a family of 2D quantum clock models and investigate the expected emergent U(1) symmetry previously studied extensively
in classical clock models. We analyze renormalization group (RG) flows versus the system size of observables computed using quantum Monte Carlo (QMC)
simulations, following methods recently developed in a study of classical models \cite{Huinew}. A schematic flow diagram is shown
in Fig.~\ref{fflows}. While we find critical behavior with exponents compatible with the 3D XY universality class, as expected \cite{Cardyt,XYe2},
we also observe an intriguing violation of the expected relationship between the exponents $\nu$ and $\nu'_q$, which govern, respectively,
the two divergent length scales $\xi$ and $\xi'$. To elucidate this surprising aspect of the quantum systems, we turn to spatially anisotropic 3D classical
XY models, where a stronger coupling in one dimension mimics the imaginary time dimension of the quantum models \cite{Cardyt}. Here we observe cross-overs between
the behavior of the isotropic model and that of the quantum models, suggesting that anisotropy qualitatively alters the RG flow of the DI perturbation in the
ordered phase. We discuss a potential role of topological line defects in this anomalous renormalization of the clock perturbation.

In the remainder of this introductory section we provide some additional background and motivations for our study. In Sec.~\ref{sub:emergent} we discuss
recent interests in emergent symmetries in quantum systems in the context of deconfined quantum critical points. In Sec.~\ref{sub:u1scaling} we describe
known facts on scaling behaviors related to emergent U(1) symmetry. In Sec.~\ref{sub:outline} we outline the organization of the rest of the paper.

\subsection{Emergent symmetries and deconfined criticality}
\label{sub:emergent}

The emergence and breaking of  U(1) symmetry has been observed in $S=1/2$ $J$-$Q$ quantum spin models on the two-dimensional (2D) square and
honeycomb lattices. These models harbor a deconfined quantum critical point separating N\'eel antiferromagnetic and VBS states breaking either
$Z_4$ (square lattice) \cite{Sandvik07,Lou09,NDQC,Jiang08} or $Z_3$ (honeycomb lattice) \cite{Pujari13,Pujari15} symmetry. The N\'eel and VBS phases
correspond to condensed and confined phases, respectively, of the deconfined spinons that exist as independent objects only at the critical point
\cite{Senthil}. The spinon confinement scale in the VBS phase is related to the U(1) length-scale $\xi'$ \cite{NDQC,Senthil2}. The most concrete
manifestation of the second length scale may be in the width of a domain wall separating domains with different VBS (or classical clock) patterns
\cite{Levin,Senthil2}. The related finite-size scaling form of the energy density of a critical domain wall exhibits puzzling differences between classical
clock models and the $J$-$Q$ model, which can be described phenomenologically with a scaling function with two relevant arguments if a certain limiting behavior
is imposed when the system size is taken to infinity \cite{NDQC}.

The emergent U(1) symmetry has also been investigated with 3D classical loop \cite{NahumX2} and dimer models \cite{CDmr}, which are also argued
to realize deconfined quantum criticality. Signs of even higher symmetries, SO(5) and O(4), of the combined N\'eel and VBS order parameters have been
observed in both the 2D quantum and 3D classical effective models \cite{Nahum,NahumX,CDmr,Suwa}, including in the surprising context of
first-order N\'eel--VBS transitions resembling spin-flop transitions in O($N$) models \cite{Zhao19,Serna,Takahashi20}. The break-down of the higher
symmetries inside the ordered phases adjacent to the deconfined critical point should also be governed by a second length scale.

\subsection{Scaling of emergent U(1) symmetry}
\label{sub:u1scaling}

The detailed form of the divergence of the symmetry cross-over scale $\xi'$ is associated with subtleties even in the
prototypical classical 3D clock models. The conventional correlation length $\xi$ and the U(1) length (which also
grows with the number of clock directions $q$ \cite{Oshikawa}) diverge as
\begin{subequations}
\label{divforms}
\begin{eqnarray}  
\xi & \sim & |t|^{-\nu},\label{xidiv}  \\
\xi_q^{\prime} & \sim &  t^{-\nu_q^{\prime}}~~~ (t>0) \label{xiprimediv},
\end{eqnarray}
where $t=T_c-T$ is the distance to the critical temperature $T_c$ and Eq.~(\ref{xiprimediv}) applies
only to the ordered phase, $t>0$, as indicated. The relationship between $\nu$ and $\nu_q^{\prime}$ follows from a two-stage
renormalization procedure \cite{Nelson,Amit}, where initially the system for small $t>0$ flows toward the U(1) symmetry breaking
Nambu-Goldstone (NG) fixed point before the clock perturbation becomes relevant and the system crosses over and begins flowing toward
the $Z_q$ breaking clock fixed point.
\end{subequations}

In an early work, Chubukov {\it et al.} already derived what turns out to be the correct exponent relationship \cite{Chubukov}, 
\begin{equation}
\nu'=\nu \left (1+\frac{|y_q|}{p} \right ),
\label{nunuprime}
\end{equation}
where $y_q < 0$ is the scaling dimension of the irrelevant clock field at the critical point and $p=2$ was determined from the
properties of the NG point. However, this result appears to have been initially largely neglected, and other relationships
were also subsequently proposed \cite{Miyashita,Oshikawa}. More recent works have also arrived at Eq.~(\ref{nunuprime}) with $p=2$
\cite{LenDel,Okubo}, but the same form with $p=3$ was argued in Ref.~\cite{LouSan}. Very recently it was shown that Eq.~(\ref{nunuprime})
also follows from a generic scaling hypothesis with two relevant scaling arguments, $tL^{1/\nu}$ and $tL^{1/\nu^{\prime}}$, but the exponent
$p$ depends on the physics of the system in a non-generic way \cite{Huinew}. A simple way to understand the exponent $p$ in finite-size
systems is that it governs the initial growth $\phi_q \sim L^p$ of a properly defined $Z_q$ order parameter $\phi_q$ [which vanishes if
the angular fluctuations of the order parameter are U(1) symmetric] for a range of system sizes $\xi < L < \xi'_q$ \cite{LouSan,Huinew}.

\subsection{Aims and paper outline}
\label{sub:outline}

As already mentioned above, emergent U(1) and even higher symmetries are of great current interest in the context of VBS
order in quantum magnets with putative deconfined quantum critical points. The U(1) symmetry emerging on the VBS side
of the transition has been observed and the exponent $\nu'$ has been extracted using different methods in $J$-$Q$ models
with SU(2) spins as well as with generalizations to SU(N) symmetry \cite{LouSan,NDQC}. The recently proposed phenomenological
scaling function \cite{Huinew} from which the relationship Eq.~(\ref{nunuprime}) can be derived offers improved ways of extracting
the exponents, including $p$, from numerical data. This motivates more detailed studies of various models with quantum phase
transitions associated with emergent symmetries.

Our goal here is to consider perhaps the simplest quantum systems with emergent U(1) symmetry---a family of 2D quantum clock models.
We find that Eq.~(\ref{nunuprime}) holds at the quantum phase transition, but with the exponent $p=3$ instead of $p=2$. Since the 2D quantum
clock model should map onto a spatially anisotropic classical 3D clock model, where the third dimension corresponds to imaginary time
in the quantum case \cite{Cardyt}, we also study classical models with varying degree of anisotropy. Here we find a cross-over behavior,
where systems with weak anisotropy initially exhibit $\phi_q \sim L^2$ scaling in the neighborhood of $T_c$ but for larger sizes cross over to
$\phi_q \sim L^3$. The cross-over length decreases when the anisotropy increases, so that for a strongly anisotropic system no clear-cut $L^2$
scaling can be observed. Conversely, when the degree of anisotropy decreases, the cross-over length increases, so that the isotropic system
only exhibits $L^2$ behavior. The cross-over behavior is also reflected in the $t$ dependence of the cross-over length, and all our results point
consistently to Eq.~(\ref{nunuprime}) with $p=3$ as the correct way to describe the asymptotic relationship between $\xi$ and $\xi^\prime_q$. While
we do not have a rigorous explanation of this surprising result, we will present a phenomenological argument hinting at the way anisotropy alters
the renormalization of the clock field at the NG point, possibly as a result of vortex line defects in the anisotropic systems.
           
The structure of the paper is as follows:
In Sec.~\ref{sec2} we review emergent symmetry in the context of the classical clock model, define the $Z_q$ order
parameter, describe the finite size scaling method used to extract the scaling dimension associated with it,
and present numerical evidence for behavior consistent with $p=2$ along with a discussion of published results supporting the same.
Detailed numerical results for contrasting behavior consistent with $p=3$ are presented for ground state phase transitions
of various quantum clock models in Sec.~\ref{sec3}. This result naturally motivates a study of the anisotropic classical clock model,
which is related to the quantum clock model through the Suzuki-Trotter formalism. This mapping is discussed in detail in Sec.~\ref{sec4}
along with our numerical evidence for qualitatively different behaviors for varying degrees of anisotropy. We present phenomenological
arguments justifying the observed behaviors in Sec.~\ref{sec4c}. Conclusions and potential future directions of research are presented
in Sec.~\ref{sec5}. Auxiliary results and the QMC algorithm developed for the quantum clock model are presented in Appendices.

\section{Dangerously Irrelevant Operator in the Classical Clock Model}
\label{sec2}

The classical 3D clock model represents a prime example of emergent U(1) symmetry. It has been studied analytically~\cite{Chubukov,Oshikawa,LenDel} and
in several numerical works ~\cite{Miyashita,Hove,LouSan,Hasenbusch11,Zhitomirsky,Okubo,Banerjee,Huinew}. The Hamiltonian is the same as the standard
ferromagnetic XY model, namely
\be\label{hcm}
H_J=-J\sum_{\langle i,j\rangle}\cos(\theta_i-\theta_j),~~~~(J>0),
\ee
with the additional constraint that $\theta_i$ is no longer a continuous angle but can only take $q$ equally spaced values,
$\{0,2\pi/q,4\pi/q,...2\pi(1-1/q)\}$. We call this the hard clock model, as the degree of freedom on a lattice site is strictly discretized. Another way of
formulating the clock model is by allowing the phase at each site to continuously vary between 0 and $2\pi$ as in the XY model, but including a
site potential of the form 
\be
H_h=-h\sum_i \cos(q\theta_i).
\ee
We call the Hamiltonian $H=H_J+H_h$ the soft clock model, and clearly the hard model is obtained from it for $h \to \infty$.
Both the hard and soft models exhibit 3D XY universality with emergent U(1) symmetry for $q \ge 5$ \cite{DIO1}. For $q=4$, the hard model maps
onto two decoupled Ising models and is different from the soft model, the latter exhibiting emergent U(1) symmetry for small values of $h$ (with the
exact bound on $h$ not known precisely \cite{Pujari15,Huinew}) while the former hosts a conventional Ising transition with no emergent higher
symmetry. From this point onwards, we only consider the hard clock model and often simply refer to it as the clock model.

\subsection{Renormalization of the clock field}

As the low temperature phase must necessarily break the discrete $Z_q$ symmetry, the corresponding operator perturbing the U(1) symmetric
XY model behaves as a DI operator \cite{Nelson,Amit} as already discussed above in Sec.~\ref{sub:u1scaling}. The scaling dimension
of the irrelevant clock operator is $\Delta_q$ and the corresponding negative scaling dimension of the field $h$ is $y_q=3-\Delta_q$ according
to standard scaling theory. Thus, at the critical point the effective field when coarse-grained at some length scale $\Lambda$ is $h\Lambda^{-|y_q|}$.
While irrelevant at the critical point, away from the critical point, inside the ordered phase, the perturbation becomes relevant, which can be
understood \cite{Amit} as a scaling correction to the irrelevant part of the form $ht \Lambda^{1/\nu'_q}$. 
This follows simply by Taylor expanding $h(\Lambda^{-|y_q|}+f(t\Lambda^{1/\nu'_q}))$ for small
$t \Lambda^{1/\nu'_q}$, where the function $f(x)$ is assumed to be analytic.
While this correction vanishes for $t \to 0$,
it eventually becomes the dominant contribution for large $\Lambda$ when $ht \not=0$. The relevant scaling correction defines the positive exponent
$\nu'_q$ in the same way as the properly scaled reduced temperature (the relevant thermal field) $t \Lambda^{1/\nu}$ is governed by the correlation
length exponent $\nu$. When set to constants, the scaled thermal and clock fields give the forms of the relevant divergent length scales already
stated in Eq.~\eqref{divforms}; $\Lambda_1 \propto \xi \sim t^{-\nu}$ is the conventional correlation length associated with the fluctuations of the
magnetization amplitude and $\Lambda_2 \propto \xi'_q \sim (ht)^{-{\nu'_q}}$ represents the length scale at which the clock perturbation becomes relevant.
Since normally the microscopic field $h$ is a constant, we can also simply write $\xi'_q \sim t^{-\nu'_q}$, though it should be kept in mind that there is
also an $h$ dependence in the overall effects of the perturbation.

It should be noted here that in many cases the bare clock field $h$ is not even easily defined, e.g., in the hard clock model considered here, where no coupling
$h$ appears explicitly but nevertheless there is some implicit strength of the local clock field (e.g., when coarse graining over a few sites). The same is
true in the case of the $J$-$Q$ quantum spin models discussed in Sec.~\ref{sub:emergent}, where the $Z_3$ or $Z_4$ symmetry of the VBS order parameter
originates from the lattice and, for a given variant of the model, there is no tunable parameter that can explicitly change the strength of the $U(1)$
symmetry breaking.

An important aspect of the problem is that the exponents $\nu$ and $\nu'_q$ are related according to Eq.~(\ref{nunuprime}) with $p=2$, where the latter
exponent is associated with the physics of the NG point. In this paper we will present evidence of $p$ changing to $p=3$ in the quantum clock models
introduced and studied in Sec.~\ref{sec3} as well as in the spatially anisotropic 3D clock model studied in Sec.~\ref{sec4}.  We here first discuss the
isotropic classical case in more detail, reviewing the Monte Carlo (MC) RG flow method of Ref.~\cite{Huinew} and also presenting some additional
numerical results demonstrating $p=2$ scaling.

\subsection{Order parameters}

The divergent length scale $\xi$ and $\xi'_q$ are both associated with order parameters.
Using the magnetization
\be
\label{mdef}
\vec{m}=\frac{1}{N}\sum_{i=1}^N \vec{m}_i,~~~ \vec{m}_i=\cos(\theta_i)\hat x + \sin(\theta_i)\hat y,
\ee
the standard Binder cumulant for an U(1) [here emergent U(1)] order parameter is defined as
\be
\label{umdef}
U_m=2-\frac{\braket{m^4}}{\braket{m^2}^2}.
\ee
The development of magnetic order can be conveniently probed using $U_m$, as it vanishes for $L \to \infty$ in the paramagnetic phase and approaches
unity when the magnetization develops a finite value with small fluctuations around this value in the ordered phase. At the critical point, $U_m$ attains
a non-trivial value between zero and unity, which is dependent on the universality class of the transition. These three fixed points allow us to probe
the response of the system to the relevant (thermal) field close to criticality.

We quantitatively analyze the emergent symmetry using a $Z_q$ order parameter defined as
\be
\phi_q=\braket{\cos(q\theta)},
\label{phiqdef}
\ee
where $\theta$ is the orientation of the global magnetization vector $\vec{m}$. In the ferromagnetic phase in the thermodynamic limit (and also for any finite
$L$ at $T=0$), $\theta$ can only take the values ${n2\pi}/{q}$, with $n$ an integer in the set $\{0,...,q-1\}$, and $\phi_q$ under this distribution of
$\theta$ evaluates to unity. In the opposite extreme limit, $\phi_q=0$ if the distribution $P(m_x,m_y)$ is circular symmetric, i.e., if it reduces to $P(m)$, 
Note that, while $\phi_q$ is not explicitly sensitive to the magnitude of the magnetization, only its angular distribution, implicitly $\phi_q$ is still
suppressed when $m$ is small in finite systems, as both the angular and amplitude fluctuations increase when the critical point is approached.

The U(1)--$Z_q$ cross-over has in some past works been investigated with an order parameter $\braket{m \cos(q\theta)}$ including the magnitude
$m$ \cite{LouSan,Zhitomirsky}. This quantity was analyzed under the assumption (which also was demonstrated analytically
in a certain limit \cite{LouSan}) that the magnitude factors out; $\braket{m \cos(q\theta)} \to \braket{m} \braket{\cos(q\theta)}$, with $\braket{m}$ obeying
the standard finite-size scaling form, $\braket{m} \sim L^{-\beta/\nu}$ ($\beta$ being the critical exponent of the magnetization below $T_c$) when $m$ is
small. While in Ref.~\cite{LouSan} this procedure for $q=4,5,6$ models delivered results for $\nu'_q$ consistent with later studies using Eq.~(\ref{phiqdef})
\cite{Okubo,Huinew}, in Ref.~\cite{Zhitomirsky} a very different result was obtained for $q=6$. The reasons for the latter discrepancy is still unclear, but
in general we advocate Eq.~(\ref{phiqdef}) as a pure $Z_q$ order parameter that is not mixed with the amplitude $m$. There are effects
of incomplete decoupling of the angular and amplitude fluctuations, and the critical scaling of $m$ governed by $\beta/\nu$ only applies when
$m$ is small. Indeed the $\braket{m} \sim L^{-\beta/\nu}$ scaling breaks down when $L > \xi$, whence $m$ is close to $m(\infty) \sim t^\beta$ in the
thermodynamic limit while the angle is still almost U(1) distributed if $L < \xi'_q$. In this regime (and for $L > \xi'_q$) the assumptions
of Refs.~\cite{LouSan,Zhitomirsky} are strongly violated \cite{note}.

\subsection{RG flows and scaling function}
\label{sub:rgflows}

Using MC results for $\phi_q$ and $U_m$, we can investigate four different types of RG flows using a diagram of the kind shown schematically
in Fig.~\ref{fflows}. Here each trajectory is for a particular value of $t=T_c-T$ and corresponds to a set of increasing system sizes. A small size marks
the start of the trajectory and increasing system size at fixed $t$ corresponds to lowering the energy scale (or increasing the coarse-graining
length scale). Pictorially, such a diagram looks very much like a standard RG flow diagram (see, e.g., Ref.~\cite{Oshikawa,Okubo}), but it should be stressed
that we are not looking at flows of couplings, but of operators conjugate to those couplings that are directly accessible in simulations.
We note here that $U_m$, being a ratio of cumulants, is not technically
conjugate to any coupling and has scaling dimension zero, but is a useful
probe to classify the phases and locate the critical point.

Two of the fixed
points in Fig.~\ref{fflows} are stable as viewed from the RG perspective---those corresponding to the paramagnetic phase ($U_m=0$, $\phi_q=0$) and the
ferromagnetic $Z_q$ breaking phase ($U_m=1$, $\phi_q=1$). The critical XY point is unstable, and since it is associated with emergent U(1) symmetry
it is located in the flow diagram at $\phi_q=0$, with the Binder cumulant taking a universal value $U_{mc}$ between $0$ and $1$. Finally,
the point $U_m=1$, $\phi_q=0$ in the diagram is the unstable NG point, where U(1) symmetry is spontaneously broken. The NG point is the stable fixed point
of the ordered phase of the 3D XY model without clock perturbation. It is never reached asymptotically in the clock model but attracts the flow to its
neighborhood if $T$ is close to $T_c$. The ultimate flow away from the NG fixed point toward the $Z_q$ point is governed by the exponent $\nu'_q>\nu$.

An actual flow diagram based on high-quality simulation data for the $q=6$ clock model was presented in Ref.~\cite{Huinew}, and various aspects of
the flow were tested to confirm the validity of an asymptotic scaling form
\be
\phi_q \sim \Phi(tL^{1/\nu},htL^{1/\nu^{\prime}_q},hL^{-|y_q|}),
\label{phiqscale1}
\ee
describing the finite-size flows with two relevant arguments and one scaling correction due to the irrelevant clock field. Since $\phi_q=0$ if $h=0$,
an expansion in the small irrelevant argument gives
\be
\phi_q \sim hL^{-|y_q|}\Phi(tL^{1/\nu},htL^{1/\nu^{\prime}_q},0),
\label{phiqscale2}
\ee
which for fixed $h$ (which has an undetermined value in the hard clock models used here) we simply write as
\be
\phi_q \sim L^{-|y_q|}\Phi(tL^{1/\nu},tL^{1/\nu^{\prime}_q}),
\label{phiqscale3}
\ee
without the proportionality constant $h$ (noting that there are also other, unknown proportionality constants). We note again that the condition
$x_2=tL^{1/\nu'} \ll x_1=tL^{1/\nu}$ can always be fulfilled, at least in principle, for large $L$, and this condition is what allows us to analyze
$\Phi(x_1,x_2)$ in three distinct limits of the arguments; 1) $x_1,x_2 \ll 1$, 2) $x_1 \gg 1, x_2 \ll 1$, and 3) $x_1 \gg 1, x_2 \gg 1$.  

Following the schematic flow diagram in Fig.~\ref{fflows} and the quantitative scaling function Eq.~(\ref{phiqscale3}) it can be seen that two
approximately scale invariant regions of the flow diagram can be identified. The standard critical scale-invariant behavior $\phi_q \sim L^{-|y_q|}$
applies when $tL^{1/\nu} \ll 1$, i.e., for $L \ll \xi$ (exemplified in Fig.~\ref{fflows} by the $T=T_c$ curve). The initial effect of $tL^{1/\nu}>0$ 
is an increase in the cumulant $U_m$, while $\phi_q$ continues to decay because of the $L^{-|y_q|}$ factor, thus steering the flow toward the NG point.
When $x_1=tL^{1/\nu}$ grows further, its effect can initially be taken into account perturbatively as an expansion of the scaling function
$\Phi(x_1,x_2=0)$, and in practice it was found that the leading effect is to cause a shallow minimum in $\phi_q$ followed by an increase \cite{Huinew}
(as we also discuss further below in Sec.~\ref{sec2e}). The value of $\phi_q$ here is still small, and in Fig.~\ref{fflows} this stage is just indicated by
the curve segment close to the horizontal axis in one of the $T<T_c$ cases. For $tL^{1/\nu}$ not small but $tL^{1/\nu'_q} \ll 1$, which corresponds to
$\xi \ll L \ll \xi'_q$, the second relevant argument in Eq.~(\ref{phiqscale3}) can still be neglected, while the first one should result in a power-law
behavior (exactly as in conventional finite-size scaling with a single relevant field \cite{Fisher72}); thus $\phi_q \sim L^{-|y_q|}(tL^{1/\nu})^a$ for
some exponent $a$ (on which we will elaborate further below). The Binder cumulant flows further toward $1$ as the relative fluctuations of $m$ diminish
with increasing system size when $L> \xi$. This second scale invariant flow can take us arbitrarily close to the NG point by choosing $t$ sufficiently
small and using large enough system sizes.

Upon further increasing $L$, when $tL^{1/\nu'}$ is no longer small, i.e., $L$ is of order $\xi'_q$ or larger, the above power laws in $L$ and $t$ still
remain valid and we can write the scaling form Eq.~(\ref{phiqscale3}) as
\be\label{pq1}
\phi_q \sim L^{-|y_q|}(tL^{1/\nu})^a g(tL^{1/\nu^{\prime}_q}),
\ee
where $g(x_2)$ is a scaling function of only the second relevant argument in Eq.~(\ref{phiqscale3}).
It was argued in Ref.~\cite{Huinew} (and supported with MC data)
that form (\ref{pq1}) captures the RG flow away from the NG fixed point all the way to the $Z_q$ fixed point (but we note that the validity of
such scaling when $\phi \to 1$ was questioned in previous work \cite{Okubo}, though supporting evidence was also seen numerically). It is useful to
recast the $L$ and $t$ dependent prefactor of $g(x_2)$ in Eq.~\eqref{pq1} more explicitly in terms of an exponent $p$ governing the size dependence,
$\phi_q\propto L^p$, when $x_2$ is still small (i.e., $g \approx 1$) and the flow is still close to the NG point. Then, for all $x_2$,
\be\label{pq2}
\phi_q \sim L^p t^{\nu(p+|y_q|)} g(tL^{1/\nu^{\prime}_q}),
\ee
and we will no longer refer to the exponent $a=\nu(p+|y_q|)$ introduced in the intermediate step in Eq.~(\ref{pq1}).

Further constraints on the above form of $\phi_q$ can be set by considering the final RG stage $\phi_q\to 1$, where the $L$ and $t$ dependence
must vanish. This necessitates $g \to (tL^{1/\nu^{\prime}_q})^b$, with the exponent $b$ chosen so that the powers of $t$ and $L$ in Eq.~(\ref{pq2}) are
canceled, which is possible only if $\nu'_q$ is constrained by the values of $p$ and $|y_q|$. These arguments result in the relationship between $\nu$
and $\nu'_q$ in Eq.~(\ref{nunuprime}), but to further determine the value of $p$ requires analysis of the physics of the NG point.

To study the RG flows with MC simulations, using small values of $t$ is necessary in order to attain the large separation in the length scales
$\xi$ and $\xi'_q$ that is required to clearly observe the second flow toward the NG fixed point, and this becomes easier for larger $q$ because
$\nu'_q/\nu$ increases with increasing $q$ \cite{Oshikawa,LouSan,Okubo,Banerjee,Huinew}. For $q=4$ (where the soft version of the model has to be
used), it is in practice  impossible to observe all the stages of the RG flow \cite{Huinew} because $y_4 \approx -0.1$ is very small, while the 
separation can be clearly achieved for $q=5$ ($y_5 \approx -1.27$) and $q=6$ ($y_6 \approx -2.5$). For larger $q$, the separation of length scales
becomes too large and it is difficult to observe the U(1)--$Z_q$ cross-over.

\subsection{Cross-over from the NG point}
\label{sec:NGcrossover}

The exponent relationship Eq.~(\ref{nunuprime}) was already some time ago demonstrated with $p=2$ by Chubukov {\it et al.}~\cite{Chubukov} (in their
Appendix B) by using the scaling form for the transverse susceptibility of the system, which should deviate from the NG form when the observed length scale (the
inverse of the momentum chosen in the susceptibility) exceeds $\xi'_q$. Later works have also justified $p=2$ \cite{Okubo,LenDel} in related ways, though
a contradictory result with $p=3$ has also been argued for \cite{LouSan}. The analysis of the MC RG flows in Ref.~\cite{Huinew} agrees with $p=2$ to a
precision of a few percent, and there should now be no doubt about this being the correct value.

We will build on the approach by Chubukov {\it et al.}~\cite{Chubukov} when we consider anisotropic systems in Sec.~\ref{sec4}. For completeness, we review
their approach leading to $p=2$ here. The starting point of the argument is that the Goldstone modes are well defined at the point where the RG flow approaches
close to the NG fixed point. The perturbative response to the discrete symmetry breaking field $h\cos(q\theta)$ can be understood through the transverse
susceptibility. Following Ref.~\cite{Chubukov}, the transverse susceptibility of a system with its global magnetization along $\theta=0$ is qualitatively
controlled by the universal function (which we motivate further in Appendix \ref{sec:app})
\be\label{stsus}
f(\bar{k},\bar{h})\propto\frac{1}{\bar{k}^2+\bar{h}}.
\ee
Here $\bar{k}$ and $\bar{h}$ are dimensionless quantities related to the physical momentum and clock field through $\bar{k}\propto k\xi$, and
$\bar{h}\propto h\xi^{y_q}$, respectively. Qualitative control of this function shifts from $\bar{k}$ to $\bar{h}$ at $\bar{k}\propto \bar{h}^{1/2}$,
i.e., when $k\xi\propto \xi^{{y_q}/{2}}$. To rewrite this relation in terms of $\nu$ and $\nu_q^{\prime}$, we only need to note that the value of $k$
where the NG susceptibility is violated corresponds to a length scale $k^{-1}$, and by definition this is exactly the U(1) cross-over length $\xi_q^{\prime}$.
Thus $\xi/\xi'_q \sim \xi^{{y_q}/{2}}$, which is exactly equivalent to the relation between the exponents $\nu$ and $\nu'$ in Eq.~\eqref{nunuprime} with
$p=2$. 

\subsection{Numerical results for the initial cross-over}\label{sec2e}

The relationship $\phi_q \sim L^pt^{\nu(p-y_q)}$, which is a reduced form of Eq.~\eqref{pq1} for $tL^{1/\nu_q^{\prime}}\to 0$ and is applicable to the
initial development of discrete order, has not been tested directly in numerics (though the consequences were probed in Refs.~\cite{Okubo,Huinew} by
analyzing data in other ways). Here, we present data from MC simulations of the $q=6$ clock model with $J=1$ and temperatures chosen to be
slightly below $T_c=2.20201(1)$ (where the number within parentheses here and henceforth is the statistical error of the preceding digit). Our method
for determining the critical point in all cases is through Binder cumulant crossing points~\cite{Binder,Luck} as in Ref.~\cite{Huinew}. We illustrate
this method for the quantum clock model in Appendix \ref{sec:app2}, where all the critical point values of the models studied in this paper are also listed.

\begin{figure}[t]
\includegraphics[width=\hsize]{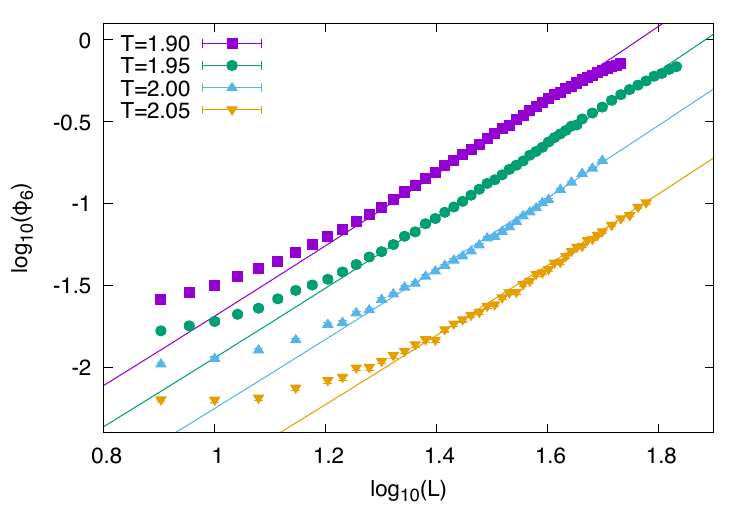}
\caption{MC results for the angular clock order parameter vs the system size for $T<T_c$ close to $T_c \approx 2.20201(1)$ of the
$6$-state clock model. The lines on this log-log plot show fits of the form $\phi_6=aL^2t^{\nu(p+|y_q|)}$ with $t=T_c-T$, $y_6=-2.509$ \cite{Banerjee},
and $p=2$. The prefactor $a$ was obtained in a fit to only the $T=1.95$ data, and the other curves were drawn using this same value of $a$.}
\label{fpiclock}
\end{figure}

Fig.~\ref{fpiclock} shows a common fit to $\phi_6$ data versus $L$ for four different choices of $T<T_c$ and using $p=2$ to test the
consistency with the expected form. We used the best known value of the irrelevant scaling dimension, $y_6=-2.509(7)$ \cite{Banerjee} (which
agrees also with the value extracted in Ref.~\cite{Huinew}, but with a  smaller error bar in the $q=6$ case). We fixed the prefactor
of the power law at $T=1.95$ and used the same factor for all other temperatures, together with the predicted scaling factor $t^{\nu(2-y_q)}$ in
Eq.~(\ref{pq2}) for the different values of $t$. We observe very good agreement between the data and the predicted form, both as regards the $t$
dependence and the $L$ dependence over a wide range of system sizes.

Here it should be noted that the $L^2$ behavior is preceded by a minimum in $\phi_6$ at a system
size $L^*(t)$, the scaling of which was studied in Ref.~\cite{Huinew}.
For general $q$, the initial growth of $\phi_q$ for $L > L^*$ arises when $tL^{1/\nu}$ becomes sufficiently
large in Eq.~(\ref{phiqscale3}) to have a significant first-order effect, i.e., the form is $\phi_q \sim L^{-|y_q|}(1+atL^{1/\nu})$ with a
positive factor $a$ such that a minimum forms. The minimum value $\phi_6$ approaches zero when $t \to 0$ as $t^{-\nu|y_q|}$ and is located at system
size $L^* \sim t^{-\nu}$. Thus, there is a crossover from the initial first-order increase of $\phi$ to the $L^2$ behavior setting in at larger
$tL^{1/\nu}$, where the perturbative expansion of the scaling function $\Phi(x_1,x_2\approx 0)$ in $x_1$ breaks down and Eq.~\eqref{pq2} applies.
The results in Fig.~\ref{fpiclock}
confirm these behaviors of Eqs.~\eqref{phiqscale3} and \eqref{pq2} with $p=2$, as well as the expectation that the overall range of system sizes
for which $\phi_6 \propto L^2$ is realized grows as $t$ decreases; with the scaling in Eq.~\eqref{pq2}, $\phi_q$ reaches an $O(1)$ value (for a
cross-over to approach its ultimate value $\phi_q=1$) at size $L \sim t^{-\nu(2+|y_q|)/2}$, which grows faster as $t \to 0$ than the also growing
size $L^* \sim t^{-\nu}$.

\section{Quantum Clock Model}\label{sec3}

The 2D quantum clock models we introduce here can be regarded as generalizations of the $S=1/2$ transverse-field Ising model (TFIM),
with a $q$-state site degree of freedom corresponding to angles $\theta_i$ as in the classical hard clock model. Quantum clock models
have garnered recent interest due to a realization of a three fold symmetric
chiral clock model in a chain of trapped
ultracold alkali atoms \cite{Qchir1D} and as a model suitable for hosting
excitations called parafermions which generalize Majorana fermions \cite{Para}.
Details of the symmetry cross-over has not been the focus of these
studies, however. Disorder effects in a 1D version of the quantum clock model
have also been studied in the context of the emergent U(1)
symmetry \cite{Senthil1D}. Here we will investigate the emergent symmetry in
2D quantum clock models, following the type of analysis developed for
the classical 3D models in the previous section.

The form of the quantum fluctuations allows for some flexibility even with a restriction to single-site off-diagonal interactions $Q_i$.
We work in the basis where the clock term is diagonal and write the Hamiltonian as
\be\label{qham}
H=-s\sum_{\braket{i,j}} \cos(\theta_i-\theta_j)-(1-s)\sum_{i=1}^N Q_i. 
\ee
For $q=2$, the clock interaction is the ferromagnetic Ising coupling of the TFIM, $-s\sigma^z_i\sigma^z_j$ when written with Pauli spin
operators, while the off-diagonal terms can be taken as a field in the $x$ direction; $Q_i=\sigma^x_i$. For $q>2$ (in practice we will consider $q=5$ and
$6$), we test three simple choices for the off-diagonal interactions to show that out results are robust to variations in the form of the quantum
fluctuations.

In practice, QMC simulations of the type we use here (described in Appendix \ref{sec:appqmc}) are restricted to models with no positive off-diagonal
matrix elements of $H$ in the chosen basis where the clock term is diagonal (otherwise we encounter the 'sign problem', where the MC weight function is
not positive definite \cite{Henelius}). Our choices are:\\

\noindent
{\bf Model (1)}: $\braket{\theta_i|Q_i|\theta_i^\prime}=\cos(\theta_i-\theta_i^\prime)+1$
only for $\theta_i-\theta_i^\prime=2\pi/q$ and zero otherwise. This choice is
most clock-like as it allows only transitions to the directions one step
away from the current direction.\\

\noindent
{\bf Model (2)}:  $\braket{\theta_i|Q_i|\theta_i^\prime}=\cos(\theta_i-\theta_i^\prime)+1$
with no constraint. This choice also provides clock-like fluctuations as it 
provides lesser weight to large changes in direction.\\

\noindent
{\bf Model (3)}:  $\braket{\theta_i|Q_i|\theta_i^\prime}=1/q$ with no constraint. This choice is a Potts-like interaction in the imaginary
time direction (as we demonstrate in Sec.~\ref{sec4a}) and contains no notion of the periodicity of the clock degree of
freedom (though overall the model still has $Z_q$ symmetry because of the diagonal term).\\

We simulate these three models using the stochastic series expansion (SSE) QMC method, generalizing an algorithm for the TFIM
\cite{SSE} with modifications to the cluster algorithm to improve efficiency. The new cluster update for $q\ge 3$ is described in
Appendix \ref{sec:appqmc}. The method delivers results free from any approximations
and can reach the low temperatures we need here to study the ground state. We extract the quantum critical points $s_c$ for models (1)-(3)
using the Binder cumulant $U_m$, Eq.~(\ref{umdef}), of the magnetization. Here $\vec m$ is diagonal in the simulations and defined exactly
as in the classical case, Eq.~(\ref{mdef}), but with the summation restricted to a single layer at fixed imaginary time. The expectation values
$\langle m^2\rangle$ and $\langle m^4\rangle$ are averaged over the time dimension and the ratio is evaluated in post processing of the
data. The inverse temperature in the simulations is taken sufficiently large ($\beta \propto L$) to ensure convergence to the ground state
of all quantities discussed here.

As expected, we find that the critical behavior matches the expected 3D XY universality class. In Appendix~\ref{sec:app2} we illustrate the determination
of the critical coupling $s_c$ and also show that the conventional critical exponents $\nu$ and $\eta$ (the anomalous dimension) are consistent with their
known 3D XY values \cite{Banerjee,3DXY}. For reference we also list the $s_c$ values for the different models. We here focus on the emergent U(1) symmetry
and the cross-over to $Z_q$ inside the ordered phase.

\begin{figure}[t]
\includegraphics[width=\hsize]{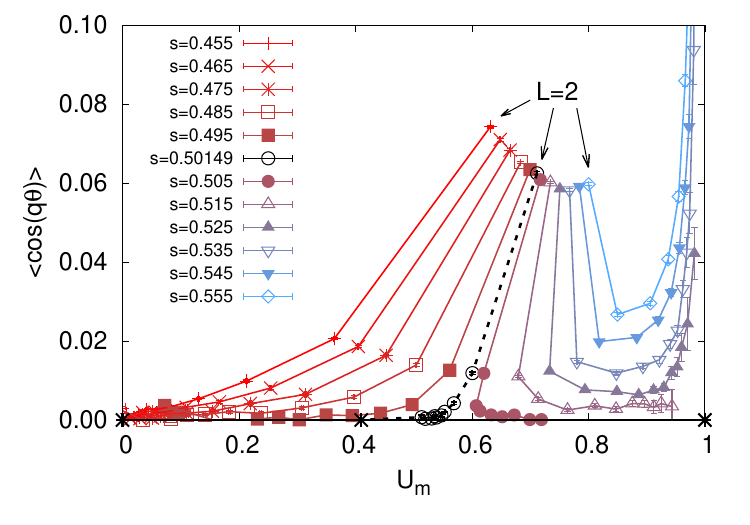}
\caption{QMC flow diagram in the space of expectation values $(U_m,\phi_q)$ for the $q=6$ quantum clock model Eq.~\eqref{qham} with kinetic terms
of type (2). The critical point is at $s_c=0.50149(6)$ (determined using the method discussed in Appendix \ref{sec:app2}). The observed flows with increasing
system size (here starting from $L=2$ and increasing the size in steps of $\Delta L=2$) for $s<s_c$ (in the paramagnetic phase phase) and $s>s_c$ (in the
ordered phase) correspond to those shown schematically in Fig.~\ref{fflows} for the classical 3D clock model with $T>T_c$ and $T<T_c$, respectively.
The three $\phi_6=0$ fixed points are marked with * and the critical separatrix between flows to the paramagnetic fixed point at $(0,0)$ and the $Z_q$
phase at $(1,1)$ is indicated with the dashed curve through the points at $s_c$. The $Z_6$ fixed point $(1,1)$ is above the upper edge of the graph.}
\label{fflow}
\end{figure}

Once again the U(1)-breaking $Z_q$ order parameter is given by $\phi_q=\langle\cos(q\theta)\rangle$, where $\theta$ is now defined using the magnetization
$\vec{m}$ of a layer at fixed imaginary time $\tau$, as explained above in the case of $U_m$, with time averaging of $\phi_q(\tau)$ also performed. We will
consider other definitions further below, with the global angle corresponding to the full space-time volume as well as a line (single spin) in the
time dimension or a line of $L$ spins within a plane.

Fig.~\ref{fflow} shows a flow diagram in the $(U_m,\phi_6)$ plane for model (2) with $q=6$. Here the separatrix, ending at $U_{mc}=0.41(1)$, between the flows
to the paramagnetic and U(1) fixed points varies slightly among the three models. It should be noted here that the critical value of the Binder cumulant is not
universal in the same strict sense as critical exponents---for a given universality class, the exact value $U_{mc}$ depends on boundary conditions and the aspect
ratio of the systems used to extrapolate to infinite size \cite{Blote,Selke}. Quantum systems have effectively adjustable aspect ratios via the choice of
the inverse temperature $\beta$, the scale of which is set by a not necessarily known velocity arising from the inherent space-time anisotropy (in the mapping to
classical systems discussed in Sec.~\ref{sec4a}). In 2D quantum spin models significant variations in $U_{mc}$ have been observed with the chosen ratio
$\beta/L$ between the inverse temperature and the system length \cite{Ma18}. Thus, the critical cumulant value should not be used to test the universality class. In
Appendix \ref{sec:app2} we show examples of critical 3D XY scaling of the order parameter of the quantum clock model. Here we focus on the U(1)--$Z_q$ symmetry
cross-over, which in Fig.~\ref{fflow} looks like the expected schematic diagram in Fig.~\ref{fflows} and the previous MC results for the classical clock model
with the same $q$ \cite{Huinew}. We proceed to investigate several aspects of the RG flow quantitatively.

\begin{figure}[t]
\includegraphics[width=\hsize]{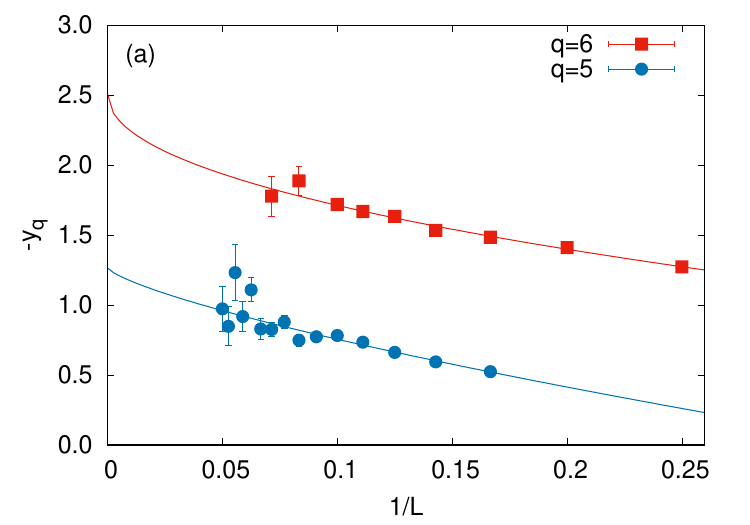}
\includegraphics[width=\hsize]{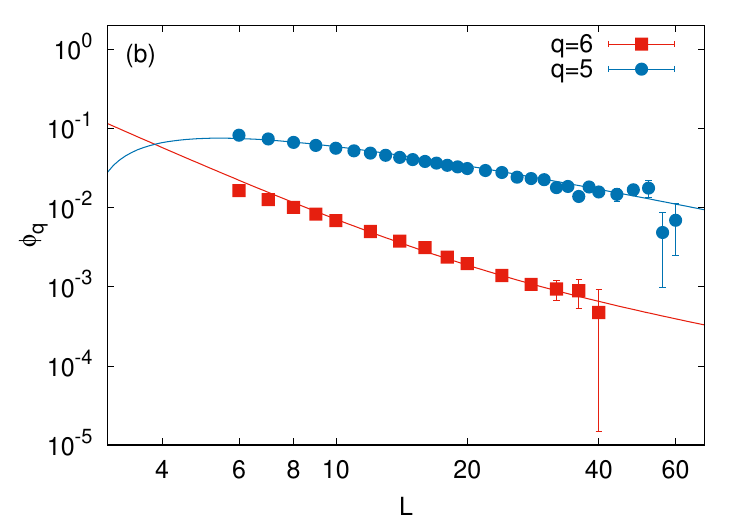}
\caption{QMC results for the $q=5$ and $q=6$ quantum clock models with off-diagonal terms of type (3).
(a) Size dependent scaling dimensions $-y_5$ and $-y_6$ calculated using Eq.~\eqref{yqldef} for $(L,2L)$ data pairs at the corresponding $U_m$ crossing
points. The curves show fits of the data vs $1/L$ to the form $-y_q(L)=a/L^b+|y_q|$,
with the constants fixed to their known scaling dimensions provided in Refs.~\cite{Banerjee,Huinew}. (b) The individual order parameters $\phi_q$ vs system size
at the $L \to \infty$ extrapolated critical points along with fit to two power laws, $\phi_q = a/L^{|y_q|}+b/L^c$, with with the same values of $|y_q|$ as in (a)
and with the correction exponent $c>|y_q|$.}
\label{fuQ6}
\end{figure}

Tests of the scaling dimensions $y_q$ of the DI operator for model (3) with $q=5$ and $6$ are shown in Fig.~\ref{fuQ6}.
This version of the model is presented as we were able to generate the highest quality data for the finite-size scaling in this case.
Fig.~\ref{fuQ6}(a) shows the flowing scaling dimension defined using system size pairs $(L,2L)$, with the $L$ dependent exponent
given (as demonstrated in Ref.~\cite{Huinew}) by $\phi_q$ data calculated either at the extrapolated infinite-size critical point or
at the flowing $(L,2L)$ cumulant crossing points (the latter of which we use here);
\be
-y_q(L) = \frac{1}{\ln(2)}\ln \left ( \frac{\phi_q(L)}{\phi_q(2L)} \right ).
\label{yqldef}
\ee
Fits to power laws in $1/L$ deliver $L\to \infty$ values in close agreement with those previously calculated using reliable methods at the 3D XY
fixed point \cite{Okubo,Banerjee,Huinew}, but with rather large error bars. In Fig.~\ref{fuQ6}(a) we therefore demonstrate consistency with the known values
(i.e., ability to obtain sound data fits) by fixing the infinite-size values and just optimizing the corrections. The decay of $\phi_q(L)$ versus
$L$ is shown explicitly in Fig.~\ref{fuQ6}(b). Here we have again carried out fits with the known values of the exponents $y_q$ fixed and included a single
finite-size correction. The inability of these fits to describe the data for the smallest system sizes indicate significant higher-order corrections that we have not
included, as we here merely wish to test the expected scaling at the higher end of the range of available system sizes.

Having established that the conventional critical behavior of the clock perturbations indeed follows 3D XY universality, we next carry out a scaling analysis of
the $q=6$ models (1) and (2) in the vicinity of the NG point where the $Z_q$ order begins to emerge. We check the
hypothesis that $\phi_q\propto L^p$ using several different definitions of the $Z_q$ order parameter. The first definition, which we already used above for
the flow diagram and scaling analysis at $s_c$, is a 2D quantity that we now denote by $\phi_{2D,q}$. 
It uses only a sum of $L^2$ spins at a fixed value of the
imaginary time $\tau$ (i.e., a 2D layer in the 3D space-time) using
$\vec{m}_i(\tau)$, where $i$ lists sites in the spatial lattice. We define
$\theta(\tau)$ by using the normalized magnetization vector defined by
$(\sum_i\vec{m}_i(\tau))/L^2$. For a given SSE configuration we
further average $\phi_q(\tau)=\cos[q\theta(\tau)]$ over many values of $\tau$
to improve the statistics.

The second version of the order parameter, denoted by
$\phi_{3D,q}$, uses all the spins in the entire SSE space-time to define the global (2+1)D magnetization as
\be
\vec{M}=\frac{1}{L^2\beta}\int_0^\beta d\tau \sum_i \vec{m}_i(\tau),
\label{m3ddef}
\ee
from which the angle $\theta$ is extracted and the order parameter calculated
with it as above.
In addition, we also study one-dimensional definitions.
For a time-like quantity $\phi_{1Dt,q}$, a coarse grained magnetization is defined for a single site $i$ by integrating the magnetization $\vec{m}_i(\tau)$
corresponding to the microscopic degree of freedom $\theta_i(\tau)$ over imaginary time as in Eq.~\eqref{m3ddef}. An order parameter $\phi_q(\theta_i)$ is
then extracted from the angle $\theta_i$ of this mean magnetization. For a similar space-like quantity, we sum the magnetization over $L$ spins on a horizontal
line or vertical line at fixed $\tau$. Translational invariance in space and time is used to average these order parameters across space-time.

\begin{figure}[t]
\includegraphics[width=\hsize]{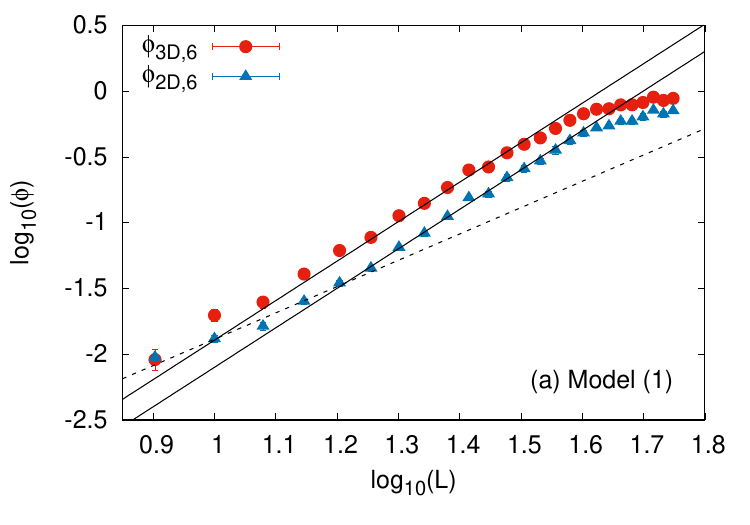}
\includegraphics[width=\hsize]{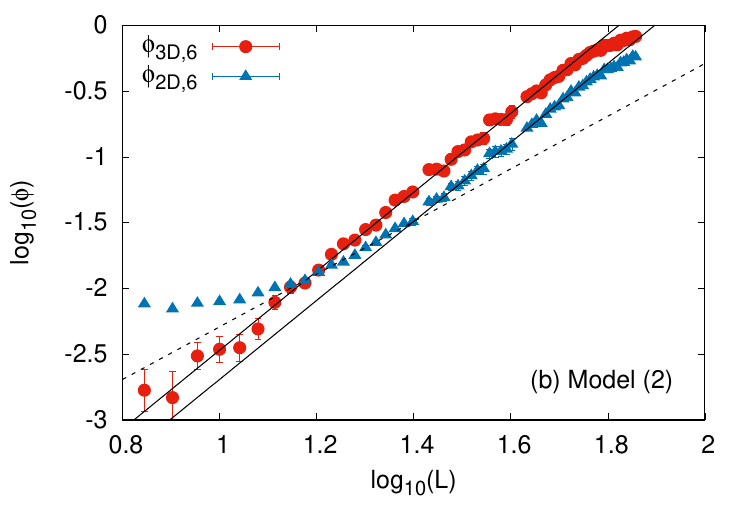}
\caption{Results for the 2D and 3D definitions of the $Z_6$ order parameter for model (1) of the quantum fluctuations in (a) and
model (2) in (b), in both cases slightly inside the ordered phase. The control parameter $t=s-s_c=0.05955$ in (a) and $t=0.0245$ in (b). The solid lines
correspond to fits in the relevant size ranges to the form $\phi_6 = aL^3$. For reference, the dashed lines show the form $L^2$.}
\label{figsp}
\end{figure}

To study the behavior of these quantities in the regime where they are small but growing with $L$, we choose values of $t=s-s_c$ such that a
large range of accessible system sizes are in the desired regime. Results for both types of order parameters and two types of quantum fluctuations
in the model are shown in Fig.~\ref{figsp}. In contrast to the classical clock model, Fig.~\ref{fpiclock}, here the $Z_q$ order parameter scales as
$L^3$ over a wide range of sizes in all cases. The cross-over from the minimum value of $\phi_6$ (discussed in connection with Fig.~\ref{fpiclock}
in Sec.~\ref{sec2e}) implies some transitional range of system sizes where the behavior is tangentially $\propto L^2$, but overall the $L^3$ behavior
appears much more plausible. It is manifested particularly well with the 3D version of the order parameter in model (2).

We next apply another technique for directly identifying the critical cross-over exponent $\nu^{\prime}_6$ and check the relationship between exponents in
Eq.~\eqref{nunuprime}, to confirm the required consequence of the unexpected $p=3$ scaling found above. As argued in Ref.~\cite{Huinew} and also supported by
the earlier results in Ref.~\cite{Okubo}, when $\phi_q\to 1$ Eq.~\eqref{pq2} can be reduced to 
\be\label{pq3}
\phi_q=1-k(tL^{1/\nu^{\prime}_q}),
\ee
where $k(x_2)$ is a function such that $k(x_2\to \infty) \to 0$. The limiting value $\phi_q=1$ was already used in Sec.~\ref{sub:rgflows}
to deduce the general exponent relation Eq.~\eqref{nunuprime}, which with $p=2$ had previously been derived \cite{Chubukov,Okubo,LenDel} by
invoking the Goldstone modes at the NG point. The form Eq.~(\ref{pq3}) is required in order to analytically connect the limits of $\phi_q \ll 1$ and $\phi_q=1$
with a single scaling function, Eq.~\eqref{pq2}. Thus, we should be able to extract $\nu^\prime_q$ from the regime where $\phi_q\approx O(1)$, and comparing the result
with Eq.~\eqref{nunuprime} should require $p=3$ based on the analysis of the size dependence in the different regime where $\phi_q \ll 1$ in
Fig.~\ref{figsp}.

\begin{figure}[t]
\includegraphics[width=\hsize]{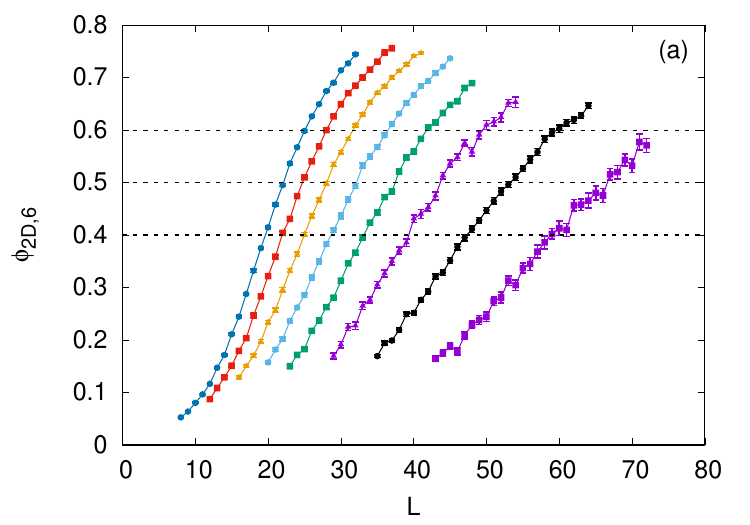}
\includegraphics[width=\hsize]{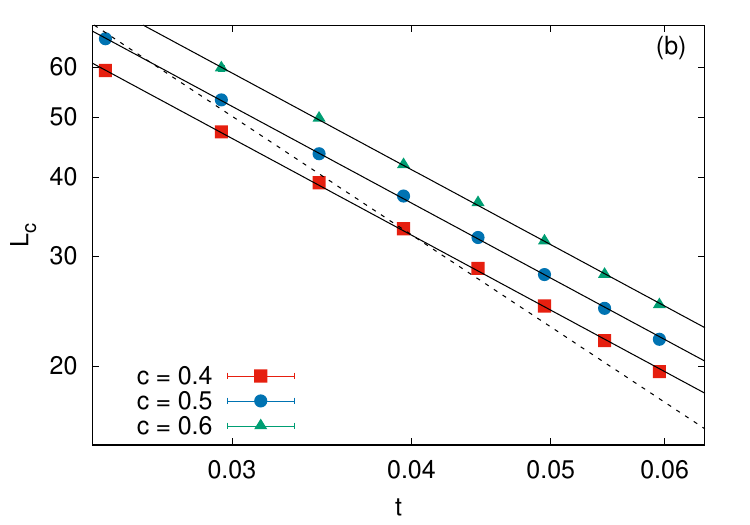}
\caption{$Z_6$ order parameter analysis for the quantum clock model with choice (2) of the quantum fluctuations.
In (a) the order parameter $\phi_{2D,6}$ is graphed vs $L$ for values of $t=s-s_c$ ranging from $t=0.0245$ (rightmost set, purple symbols) to
$t=0.0595$ (leftmost set, blue symbols). Polynomial fits to these data give sizes $L_c(t)$ at which the order parameter takes constant values $c$. 
Such $L_c$ values extracted for $c=0.4$, $0.5$, and $0.6$ are graphed vs $t$ in (b). The solid lines show fits to the form $L_c=at^{-\alpha_3}$, with 
$\alpha_3$ the exponent $\nu'_6$ predicted from Eq.~(\ref{nunuprime}) with the anomalous NG exponent $p=3$, and the dashed line is a fit with $\alpha_2$}
\label{figd2p}
\end{figure}

One can proceed in different ways to extract $\nu'_q$ using Eq.~(\ref{pq3}), e.g., by optimizing data collapse when graphing $\phi_q(x)$
versus $x=tL^{1/\nu^{\prime}_q}$ \cite{Okubo}. Here we follow the approach of Ref.~\cite{Huinew}, where a constant $c$ sets a horizontal cut $\phi_q=c$ in
the flow of $\phi_q(L)$ with $L$ at fixed $t$. The cut for a given $t$ defines a system size $L_c(t)$, as illustrated in Fig.~\ref{figd2p}(a).
Assuming the scaling form $\phi_q=f(x)$ according to Eq.~(\ref{pq3}), we can Taylor expand $f(x)$ about any value of the control parameter $x=x_0$, with a
result that we can simply write as $\phi_q=a+b(x-x_0)$ with some constants $a$ and $b$. In principle we can choose $x_0$ such that $\phi(x_0)=c$, and then
$bx=b_0$ with $b_0=bx_0+c-a$, from which we obtain
\be
L_c(t)=A_ct^{-\nu_q'},
\label{loglstar}
\ee
with some constant $A_c$ that depends on the choice of cut value $c$. Thus, we can extract $\nu'_q$ by fitting a power law to a set of interpolated
(i.e., not restricted to integers) system sizes $\{L_c(t)\}$ for different values of $t$. The form Eq.~(\ref{loglstar}) clearly just reflects the
assumption that $\phi_q(t,L)$ is a function of only $tL^{1/\nu^{\prime}_q}$, which in practice
of course should be expected to only hold asymptotically for large $L_c$ and small $t$. Deviations from the scaling form will be reflected as corrections
to Eq.~(\ref{loglstar}). If the corrections are small for system sizes and $t$ values that can be reached in practice, we can obtain a reliable estimate
for the value of $\nu_q^{\prime}$ (with scaling corrections included in the analysis if needed).

Given the $\nu,\nu'$ exponent relation in Eq.~\eqref{nunuprime},
the known 3D XY exponent $\nu=0.6717(1)$ \cite{3DXY}, and the scaling dimension $y_{6}=-2.509(7)$ of the $q=6$ clock field,
we can predict the exponent $\nu'_6$ for the expected NG exponent $p=2$ and also for the anomalous value $p=3$ that we found above. We denote by $\alpha_p$
the expected value of $\nu'_6$ corresponding to a given value of $p$; with the above XY exponents we have  $\alpha_2=1.514(1)$ and $\alpha_3=1.234(1)$.
We test the value of $\nu'_q$ this way for quantum clock model (2), using both the 2D and 3D definitions of the $Z_q$ order parameter, which should
be expected to exhibit the same scaling behavior.

\begin{figure}[t]
\includegraphics[width=\hsize]{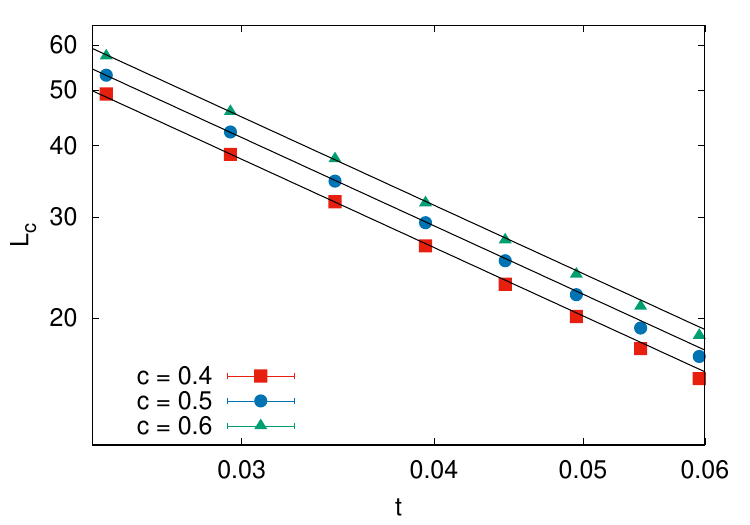}
\caption{The same kind of analysis as in Fig.~\ref{figd2p} but for the 3D version of the $Z_6$ order parameter with quantum fluctuations of type (2).
The interpolated  $L_c$ values vs $t$ for three cuts $\phi_{3D,6}=c$ have been fit (lines shown) to the form
$L_c=at^{-\alpha_3}$, with the predicted value of $\alpha_3$.}
\label{figd3p}
\end{figure}

System sizes $L_c(t)$ for several values of $t$ are extracted from the $Z_6$ order parameter calculated for a series of system sizes
in Fig.~\ref{figd2p}(a), using the 2D order parameter $\phi_{2D,6}$ and cut lines at $c=0.4,0.5,0.6$. We show a log-log plot of the so
obtained points $L_c(t)$ in Fig.~\ref{figd2p}(b). For all three values of $c$, we observe behaviors fully consistent with the power law
predicted when $p=3$ (while $p=2$ fits cannot describe the data at all). We carried out the same analysis for the 3D version
$\phi_{3D,6}$ of the order parameter and graph results along with $p=3$ fits in Fig.~\ref{figd3p}. Except for the points for the largest
$t$ values graphed, these fits work well and the trends in all data sets indicate small, rapidly decaying scaling corrections to the
expected $p=3$ form as $t$ decreases. Thus, in the case of version (2) of the quantum clock model, we have shown consistent $p=3$ scaling of
$\phi_6$ both in the initial growth form versus $L$ and in the $t$ dependence of the length scale $L_c$ when $\phi_6$ approaches unity as
the $Z_q$ fixed point is approached.

\begin{figure}[t]
\includegraphics[width=\hsize]{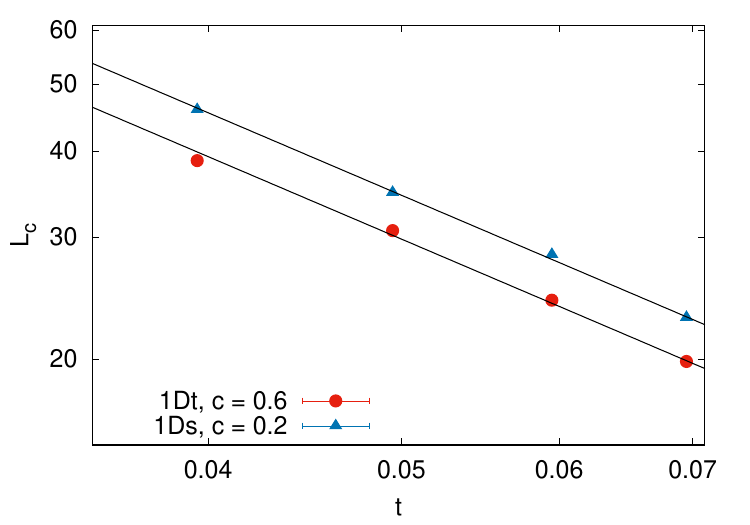}
\caption{Scaling analysis with the line versions of the $Z_6$ order parameter (spatial $1Ds$ and along imaginary time $1Dt$) for quantum fluctuations of type (2).
The system sizes $L_c$ are extracted as in Fig.~\ref{figd2p}(a). Fits are shown to the form
$L_c=at^{-\alpha_3}$ with the predicted $p=3$ exponent.}
\label{figd1p}
\end{figure}

To confirm that this behavior is truly isotropic in space-time, we also check the scaling using the line definitions of the
$Z_6$ order parameter. As shown in Fig.~\ref{figd1p}, we find the same $p=3$ governed growth of $L_c$ for both the space- and time-like
line order parameters. Here we have chosen the largest possible value of $c$ for which the size $L_c$ can be consistently calculated
using the available computer resources; $c=0.6$ for $\phi_{1Dt,6}$ and $c=0.2$ for $\phi_{1Ds,6}$.

\begin{figure}[t]
\includegraphics[width=\hsize]{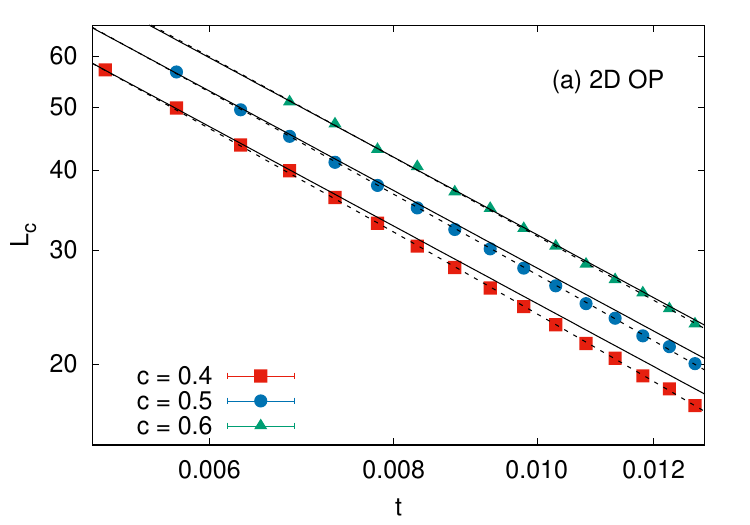}
\includegraphics[width=\hsize]{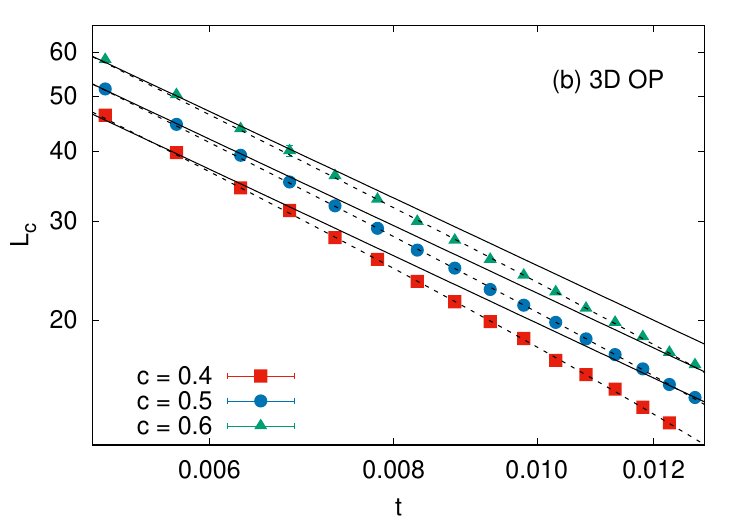}
\caption{Scaling analysis of the 2D (a) and 3D (b) $Z_6$ order parameters for the clock model with choice (3) for the quantum fluctuations. The $t$ dependence
of $L_c$ is shown for three different values of the cut parameter $c$, as in Fig.~\ref{figsp}(b) and Fig.~\ref{figd2p}. The solid lines show fits to the
form $L_c=at^{-\alpha_3}$ with the predicted $p=3$ exponent and adjustable $a$, while the dashed lines include a correction $b$; $L_c=at^{-\alpha_3}+b$.}
\label{figd23}
\end{figure}

To further test the universality of the $p=3$ scaling, we next turn to clock model (3) and carry out the same kind of analysis.
As shown in Fig.~\ref{figd23} for both the 2D and 3D order parameters, the divergence of $L_c$ with decreasing $t$ can also in this
case be fitted to the power law with exponent $\alpha_3$. In this model we also observe significant scaling corrections, and the fit
to a single power law only works for the smallest values of $t$. In Fig.~\ref{figd23} we also show fits with a constant added as
a scaling correction, which results in good fits for all the $t$ values considered. The correction is most likely a second power
law with exponent close to $0$, but the fits shown (where the correction exponent is $0$) are already close to optimal. Here it
is also important to note that the expected behavior with exponent $\alpha_2$ represents a faster divergence than exhibited by the data
for the smaller $t$ values, and the observed scaling corrections actually lead to an effective exponent (defined locally for some range of $t$)
that moves further away from $\alpha_2$ and toward $\alpha_3$ as $t$ decreases. We find it unlikely that there would be yet another
cross-over back to $p=2$ behavior for smaller values of $t$ than we have been able to reach here.

The results presented here lead to the conclusion that the U(1)--$Z_q$ cross-over is governed by a new exponent $p=3$ close
to the NG point, and this is a universal feature independent of the particular form of the kinetic terms in the quantum clock model.
The consistency of the scaling behaviors observed with different (1D, 2D and 3D) definitions of the $Z_q$ order parameter further
supports this conclusion.

\section{Anisotropic classical clock models}
\label{sec4}

As the main purpose of this section is to better understand the numerical results for
the 2D quantum clock model presented in Sec.~\ref{sec3}, we first demonstrate the mapping of
the quantum system to a 3D anisotropic clock model using the Suzuki-Trotter formalism in
Sec.~\ref{sec4a}. In Sec.~\ref{sec4b} we present numerical results supporting a $p=2$ to $p=3$
cross-over and a generalization of the scaling form Eq.~\eqref{pq2} of the $\phi_q$ order
parameter 

\subsection{Quantum to Classical Mapping}\label{sec4a}

We here carry out the mapping of the $d=2$ dimensional quantum system to an effective classical system in $d+1=3$ dimensions based on the
transfer matrix representation of the partition function $Z={\rm Tr}[\exp(-\beta H)]$ \cite{Suzuki71}. We closely follow the treatment of the TFIM,
as discussed e.g., by Cardy \cite{Cardyt}, but the more complex clock model brings some new aspects of how to exactly match the details of the quantum
and classical models.

We can construct a quantum clock model
in $d=2$ spatial dimensions using a Hamiltonian of the form $H=V+K$, where $V$ is diagonal in the basis of a clock degree of freedom with
$q$ states on each site and $K$ acts as the kinetic term, generating quantum fluctuations through off-diagonal matrix elements connecting
different clock states, as in Eq.~(\ref{qham}). With $[V,K]\neq 0$, the Suzuki-Trotter expansion \cite{Suzuki71} allows us to approximate the partition 
function using
\be
e^{-\beta(V+K)}=
\lim_{n\to\infty}(e^{-\frac{\beta}{n}V}e^{-\frac{\beta}{n}K})^n.
\ee
It can be shown that this expression for finite $n$ such that $\Delta\tau=\beta/n \ll 1$ leads to an error $O(\Delta\tau^2)$ in $Z$ and most physical 
observables \cite{Suzuki71,SZref}. As we are concerned with the ground state behavior for finite size systems, we take $\beta\to \infty$ or scale $\beta$ 
as $L^z$, where $z=1$ in our case (which is also demonstrated by our mapping). We consider $\Delta\tau \ll 1$ and at the end can let
$\Delta \tau \to 0$ for an exact mapping.

By using the quantum transfer matrix
\be\label{qT}
T=e^{-\Delta\tau V}e^{-\Delta\tau K},
\ee
we can connect $Z={\rm Tr}[T^n]$ to the partition function for a 3D classical clock model,
by interpreting $T$ as a classical transfer matrix for a system with $n$ layers, each layer having $N=L^2$ spins. In this interpretation as
a 3D classical anisotropic model, $V$ contains all the intraplane interactions and $K$ contains only interplane interactions.

To create an equivalence to a specific classical model, we now consider the actual classical transfer matrix $T_c$ corresponding to an anisotropic
classical clock model. We write this $q^N\times q^N$ transfer matrix as a product $T_c=V_cK_c$ of intra- and interplane matrices and determine
their contents to match Eq.~(\ref{qT}).

The inplane matrix $V_c$ is diagonal, with the non-zero matrix elements given by
\be
[V_c]_{l,l}=e^{-\beta_c H_{V}(C_l)},
\ee
where $\beta_c$ is the inverse temperature of the classical model and $C_l$ corresponds to configuration number $l \in \{1,\ldots ,q^N\}$ of the spins
in a layer. The  correspondence to the quantum case, as seen in Eq.~\eqref{qT}, requires us to choose  $V_c=\exp(-\Delta\tau V)$. As 
$V$ is diagonal in the chosen basis of the quantum system, the in-plane Hamiltonian for the classical equivalent model can be written as
\be\label{Hip}
H_{V}=-\sum_{\langle i,j\rangle_\parallel}J_\parallel\cos(\theta_i-\theta_j),
\ee
where the summation is over nearest-neighbor sites within a plane. The corresponding coupling must be taken as 
\be\label{jv}
\beta_c J_\parallel=s\Delta\tau,
\ee
where $s$ is the diagonal coupling constant of the quantum clock model as defined in Eq.~(\ref{qham}).

The correspondence is more involved for the matrices $K_c$ and $K$ describing the imaginary time direction.
$K_c$ is an off-diagonal matrix with elements given by
\be
[K_c]_{l,m}=\exp\{-\beta_c H_{K}(C_l,C_m)\},
\ee
where we here aim for $H_K$ to be of the same form as $H_V$ and therefore assign the interlayer Hamiltonian 
\be\label{Hop}
H_{K}=-\sum_{\langle i,j\rangle_\perp}J_\perp[\cos(\theta_i-\theta_j)-1],
\ee
where the index $K$ denotes couplings only between sites in two different (adjacent) layers and a global shift has been added in order to 
eliminate the diagonal matrix elements. By choosing this specific clock form of $H_K$, we restrict the form of the kinetic term in the quantum model
for simplicity here. To make a direct comparison to $\exp(-\Delta\tau K)$ of the quantum system, we use $\Delta\tau \ll 1$ to expand 
$\exp(-\Delta\tau K)$ as $I-\Delta\tau K$, where $I$ is the identity matrix. To compare $K_c$ and $I-\Delta\tau K$, we need to consider 
only a reduced single-site $q\times q$ matrix $K_c^r$ ($K^r$) for each site, and the full operator is just the tensor product of 
$L^2$ copies of such a matrix. To build the analogy between the quantum and classical models, we consider the off-diagonal matrix element 
between states $u$ and $v$ with angular difference $\delta\theta \in \{2\pi/q,...,2\pi(1-1/q)\}$ (i.e., $u \not= v$).
Comparing $I-\Delta\tau K^r$ and $K_c^r$ leads to
\be
-\Delta\tau(1-s)[K^r]_{u,v}=\exp\{\beta_c J_\perp[\cos(\delta\theta)-1]\}.
\ee
where $1-s$ is the overall factor of the off-diagonal term in the definition of the quantum clock model in Eq.~(\ref{qham}).
Setting $[K^r]_{u,v}<0$, corresponding to the sign of the kinetic terms in our quantum clock Hamiltonian, and noting that the
argument of the exponential is necessarily negative, we see that
\be\label{jk}
\beta_c J_\perp(\delta\theta)=\frac{|\ln|\Delta\tau (1-s)[K^r]_{u,v}||}{1-\cos(\delta\theta)}.
\ee
Thus, the classical clock model with interplane Hamiltonian Eq.~(\ref{Hop}) does not correspond exactly to any of the choices (1)-(3) of kinetic
term in our quantum clock model specified after the general form Eq.~\eqref{qham},
but we could in principle adapt $H_K$ to match our choices. However, the detailed
form of the classical Hamiltonian should not play any role in the universal scaling properties. Our different quantum models also correspond to a wide range of
different anisotropic models, ranging from the ``single clock step'' in model (1), where only $\delta\theta=2\pi/q$ is possible, to the Potts-like
version (3), where all transitions have the same weight. 

Note that, up to a multiplicative logarithm, the overall coupling ratio $J_\perp/J_\parallel$ diverges as $1/\Delta\tau$, independently of the classical inverse
temperature $\beta_c$ (which cancels out in the ratio). The quantum model therefore corresponds to the extreme anisotropic limit of the classical model, as expected
\cite{Cardyt}. An important aspect of Eq.~\eqref{jk} to note is that increasing the quantum fluctuation parameterized by $[K^r]_{u,v}$ {\it reduces} the
corresponding classical coupling $J_\perp(\delta\theta)$, under the necessary condition $\Delta\tau (1-s)[K^r]_{u,v} \ll 1$ for the Suzuki-Trotter approximation to
properly reproduce the quantum mechanics of the system. We also note that, in classical systems the anisotropy ratio is normally held constant, and $\beta_c$
is used to drive the system through its thermal phase transition. In the quantum model, the tuning parameter $s$ appears in front of both terms
in the Hamiltonian Eq.~(\ref{qham}). In the corresponding classical model, not only does the effective value of $\beta_c$ [the overall magnitude of both couplings
in Eqs.~(\ref{jv}) and (\ref{jk}), which can be absorbed into $\beta_c$] change with $s$, but the ratio $J_\perp/J_\parallel$ also changes. Thus, the way
the critical point is approached differs between the two cases, but this difference is inconsequential when studying critical scaling properties.

Even though the quantum system formally corresponds to the extreme anisotropy limit of the classical model, the physical anisotropy in the
quantum model is still finite because a given propagation time $\tau$ corresponds to a divergent number, $\tau/\Delta\tau$, of time slices.
Therefore, the velocity of excitations determining the physical space-time anisotropy is finite. In the effective classical system, even an
extreme anisotropy is irrelevant in the conventional RG sense \cite{Cardyt,XYe2} and the critical fixed point remains that of the isotropic
3D XY model. Our results for the quantum clock model in Sec.~\ref{sec3} nevertheless suggest that the anisotropy qualitatively affects the response
of the system to the clock perturbation.
The aim of the analysis of the classical anisotropic model following below is to systematically investigate the scaling of the clock
order parameter $\phi_q$ with increasing anisotropy, to connect to the quantum limit and test whether the new $p=3$ scaling behavior found in
Sec.~\ref{sec3} is indeed realized as a consequence of anisotropy. As we will see, even finite anisotropy appears to be relevant in changing the
nature of the U(1)--$Z_q$ cross-over.

In the analysis above, the number of time slices $n$ is also the number of classical layers, which, thus, should be proportional to the inverse 
temperature $\beta$ of the quantum system (for a fixed small $\Delta\tau$). Though formally the mapping between the classical and quantum models
can become exact only in the limit of infinite anisotropy, in practice the modification of the interactions if $\Delta\tau$ is very small but not
zero are irrelevant. Thus, we expect that the quantum model also eventually renormalizes to the isotropic 3D XY fixed point, as we have
also confirmed numerically in Sec.~\ref{sec3} (though with the different symmetry cross-over, which will be further elucidated below). As
the dynamic exponent $z=1$ for the critical quantum system, in our calculations we need to increase $\beta$ with the system size at least as
$\beta \propto L$ (as we did in our simulations in Sec.~\ref{sec3}) and the asymptotic quantum-critical scaling for $L \to \infty$ is independent
of the proportionality factor chosen. The QMC simulations in Sec.~\ref{sec3} are carried out directly in the limit $n \to \infty$, since the
SSE \cite{SSE,Sandvik10} method (with implementation for the quantum clock models discussed in Appendix \ref{sec:appqmc}) is based on an
exact representation of the imaginary-time continuum. In the classical systems studied below,
we will use $n=L$.

\subsection{Monte Carlo Results}\label{sec4b}

Following the analysis developed above, we define an anisotropic clock
model on a cubic system of linear size $L$, with the following Hamiltonian:
\be
H=-J_{\parallel}\sum_{\langle i,j\rangle_\parallel}\cos(\theta_i-\theta_j)
-J_{\bot}\sum_{\langle i,j\rangle_\bot}\cos(\theta_i-\theta_j),
\ee
where $\parallel$ ($\bot$) denotes nearest-neighbor bonds $\langle i,j\rangle$ in directions corresponding to
space (imaginary time). As the quantum model is expected to correspond to $J_{\bot} \gg J_{\parallel}$, we define both
coupling constants using an anisotropy parameter $\lambda\in [0,1)$:
\be
J_{\bot}=1+\lambda,~~~~J_{\parallel}=1-\lambda.
\ee

\begin{figure}[t]
\includegraphics[width=\hsize]{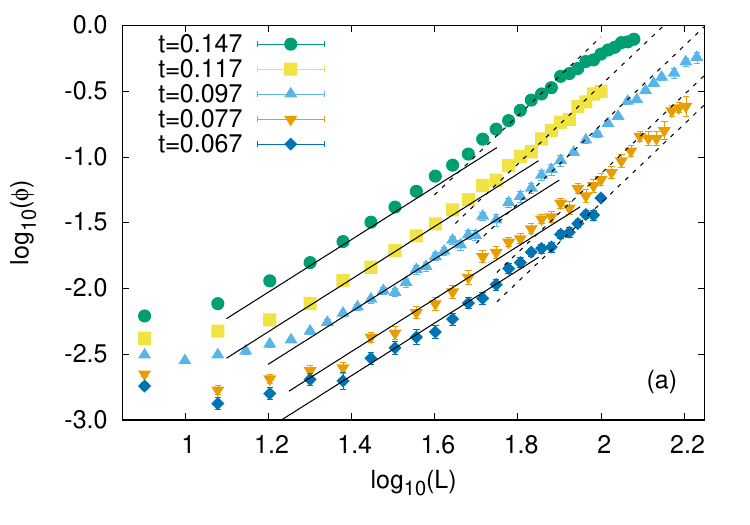}
\includegraphics[width=\hsize]{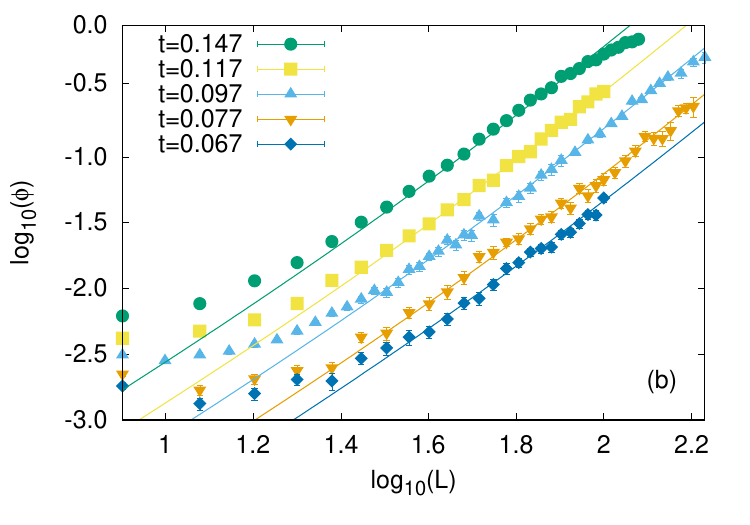}  
\caption{Scaling analysis of $\phi_6$ as a function of system size for the anisotropic clock model with $\lambda=0.5$
and several values of $t=T_c-T$, shown in log-log plots. (a) Fits to the form $\phi_6=a_pL^pt^{\nu(p-y_q)}$, expected from
Eq.~\eqref{pq2} when $g \approx 1$, for $p=2$ (solid lines) and $p=3$ (dashed lines). The factors
$a_p$ were determined for $t=0.097$ using appropriate size ranges, and the functions are plotted for
all values of $t$ with the same constants $a_2$ and $a_3$. (b) The same data as in (a) fitted to
the cross-over form Eq.~(\ref{pq2new}), where the fitted constants are $c_2=0.008(1)$ and
$c_3=0.0005(1)$.}
\label{fpanp5}
\end{figure}

In order to systematically investigate the effects of anisotropy we study
a wide range of values of $\lambda$. Studies such as these of the critical behavior have
been carried out numerically for the Ising model~\cite{RGIsingN},
where it was confirmed that the low-energy universal physics at the phase transition is not
affected by the strength of the anisotropy. There is no reason to doubt the irrelevance
of the anisotropy in the conventional critical behavior of the XY model, and we indeed
find consistency with 3D XY exponents (which we will not discuss here), including the
$Z_q$ scaling dimensions $y_q$. However, as we will see, there are nevertheless qualitative
effects of anisotropy on the emergent U(1) symmetry. We focus here on the case $q=6$ and
study the behavior at $T<T_c$ (with the $T_c$ values listed for all cases in
Appendix \ref{sec:app2}).

\begin{figure}[t]
\includegraphics[width=\hsize]{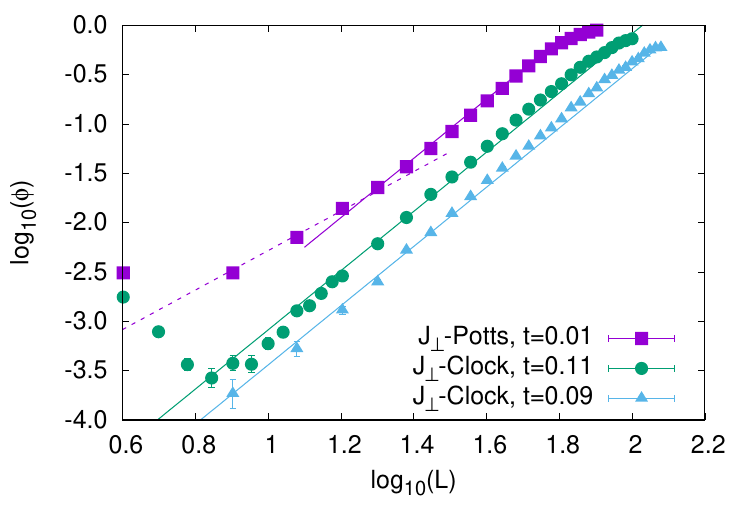}
\caption{Scaling of $\phi_6$ as a function of system size for the anisotropic clock model with $\lambda=0.9$
and two values of $t$, shown on a log-log scale. Also shown are results for one value of $t$ for the model
with Potts-like interaction, Eq.~\eqref{htpotts}, in the imaginary time direction. Common fits of the same type as
in Fig.~\ref{fpanp5} were carried out with the size exponent $p=3$ in the case of the clock-type coupling $J_{\bot}$ 
in the ``time'' dimension (blue triangles and green circles). For the system with
Potts-like $J_{\bot}$ coupling (purple squares) fits with both $p=2$ (dashed line) and $p=3$ (solid line)
are shown. Note that the $t$ values for the two types of imaginary time couplings are not directly comparable,
as the effect of changing $t$ depends on the type of interaction.} 
\label{fpanp9}
\end{figure}

We begin by studying the initial growth of the order parameter $\phi_6$, as we did in
Fig.~\ref{fpiclock} in the isotropic case. As shown in Fig.~\ref{fpanp5}(a), at $\lambda=0.5$
we find approximately $\phi_6 \propto L^2$ behavior after the cross-over from the minimum
value of $\phi_6$ (discussed in the isotropic case in Sec.~\ref{sec2e}). However, the results are not
as convincing as in the isotropic case, with what appears to be a cross-over to yet another form of
the $L$ dependence before the final approach to $\phi_6=1$. The intermediate behavior can be well
fitted to the predicted general form with $p=3$ in Eq.~(\ref{pq2}), as we found in the case of the quantum
clock model in Sec.~\ref{sec3}.

\begin{figure}[t]
\includegraphics[width=\hsize]{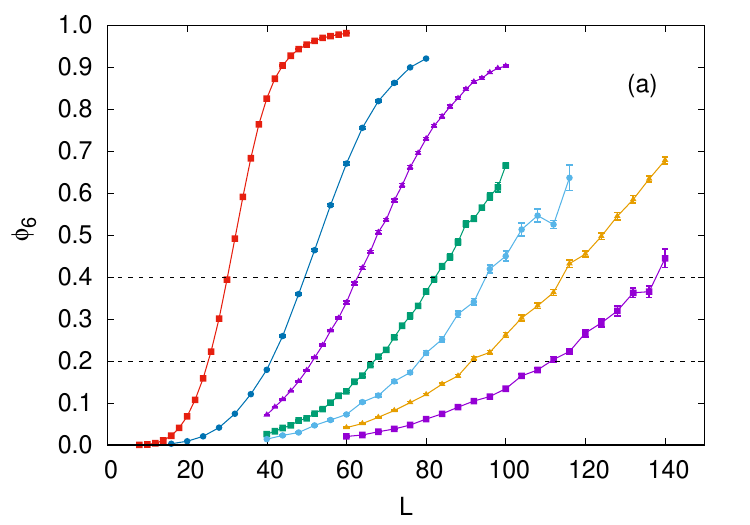}
\includegraphics[width=\hsize]{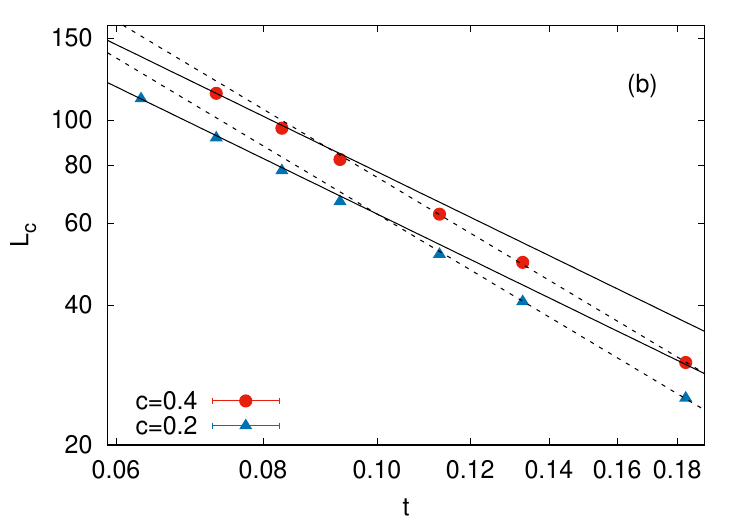}
\caption{Results for anisotropy $\lambda=0.95$.
(a) $\phi_6$ vs system size for $t$ ranging from $t=0.063$ (rightmost set, purple symbols) to $t=0.183$ (leftmost set, red symbols).
The lines shown at $\phi_6=0.2$ and $0.4$ correspond to the cuts denoted $c$ in the text, for which a corresponding size $L_c$ (and its
statistical error) is defined such that $\phi_6(L_c)=c$ is obtained by polynomial interpolation based on $\approx 10$ data points.
(b) Extracted $L_c$ values graphed vs $t$ for the two cut values. The lines show the form $L_c=at^{-\alpha_p}$ with $p=2$ (dashed lines) and
$p=3$ (solid lines). The constant $a$ was adjusted to fit the data points for the larger ($p=2$) and smaller ($p=3$) values of $t$.}
\label{figda0p95}
\end{figure}

It is important to note here that Eq.~\eqref{pq2} automatically sets the
exponent governing the $t$ dependence once a value of $p$ is chosen. This implies that individual fits should not
be performed for the different choices of $t$, and only a common non-universal prefactor should be optimized as we
already did for the isotropic model in Fig.~\ref{fpiclock}. This is also the technique used in Fig.~\ref{fpanp5}(a),
where the central data set ($t=0.097$) was first fitted to extract the overall proportionality constants $c_2$ $(c_3)$ for
$p=2$ $(3)$, and these constants together with the $t$ dependence in Eq.~\eqref{pq2} were then applied for all data sets.
The good agreement to several data sets for both the $p=2$ and $p=3$ fits  strengthens the conclusion that Eq.~\eqref{pq2}
predicts the $L$ and $t$ dependence.

If indeed we have different regions of $p=2$ and $p=3$, the validity of the $t$ dependence in both cases suggests
a cross-over originating from the argument $x_1=tL^{1/\nu}$ of the function $\Phi(x_1,x_2)$ in Eq.~(\ref{phiqscale3}).
Instead of the power law $x_1^a$ generated when the argument is large in Eq.~\eqref{pq1}, our numerical results strongly
suggest a sum of two different power laws, so that Eq.~(\ref{pq2}) is replaced by (specifically
for the powers $p=2$ and $p=3$ found here):
\be\label{pq2new}
\phi_q \sim (c_2 L^2t^{\nu(2+|y_q|)}+c_3 L^3t^{\nu(3+|y_q|)})g(tL^{1/\nu^{\prime}_q}).
\ee
With this form, which still conforms to the general scaling form $\Phi(x_1,x_2)$ in Eq.~\eqref{phiqscale3},
dominant $L^2$ behavior is always guaranteed for some range of $L$ if $t$ is sufficiently small, while
for any $t$ the $L^3$ behavior eventually takes over and persists until $L \sim \xi'_q$ and $\phi_q \to 1$.

In Fig.~\ref{fpanp5}(b) we show a fit of the above
form to all our $\lambda=0.5$ data, assuming $g(x_2)=1$ in the relevant range of $(t,L)$. While the fits are not perfect for
the smaller sizes, as expected because of the cross-over behavior from the region of the $\phi_q$ minimum
(as in the case of the isotropic model, Fig.~\ref{fpiclock}), they clearly describe the $p=2$ to $p=3$  cross-over aspects
of the $L$ dependence. Indeed, where the fits represent the data well, the behavior corresponds to the regime where neither of
the terms dominate, as seen by comparing Figs.~\ref{fpanp5}(a)  and \ref{fpanp5}(b), and the good fits persist until the
cross-over when $\phi_6$ approaches $1$. We will discuss phenomenological arguments for $p=3$ asymptotics
and the cross-over behavior in Sec.~\ref{sec4c}, after providing more evidence for the existence of $p=3$ scaling
for stronger anisotropy.

The $p=3$ behavior is seen more clearly at $\lambda=0.9$, as shown in Fig.~\ref{fpanp9},
where $L^2$ behavior cannot even be observed for any of the $t$ values considered in the model
with the same type of clock interaction in all directions (the blue triangles and green circles).
An approximate $L^3$ behavior applies in these systems over a full order of magnitude of system sizes.

It is also important here to consider other forms of anisotropy, as the quantum to classical mapping in Sec.~\ref{sec4a}
does not create exactly the same term in the imaginary time direction as the simplest clock model with only
$J_\parallel \not= J_\perp$. A Potts-like temporal interaction,
\be\label{htpotts}
H_t=-J_{\bot}\sum_{\langle i,j\rangle_\bot} (\delta_{\theta_i,\theta_j}-1),
\ee
is of special importance, as it can be mapped faithfully into the quantum case by following the derivations in
Sec.~\ref{sec4a}, where the modified version of Eq.~\eqref{jk} will lack the term $\cos{\delta\theta}$ in the denominator.
For this model, as shown in Fig.~\ref{fpanp9} for $\lambda=0.9$ (purple squares), we observe a cross-over from $L^2$ to
$L^3$ behavior, similar to Fig.~\ref{fpanp5} for the conventional anisotropy of strength $\lambda=0.5$. These results show
that the $L^2$ to $L^3$ cross-over behavior is robust to the kind of interaction used in the imaginary time direction and
also suggest that the range of validity of the $\phi_q \sim L^3$ scaling increases for higher anisotropy.

We complete the study of the classical model with a direct evaluation of $\nu'_q$ using the size intercept method discussed for the
quantum clock model in Sec.~\ref{sec3}. As shown for $\lambda=0.95$ in Fig.~\ref{figda0p95}(b), $L_c(t)$ points extracted by
interpolating within the data sets in Fig.~\ref{figda0p95}(a) for two different values of the cut $\phi_6=c$ appear to both
tend to the expected power law with exponent $\alpha_3$ (i.e., the predicted value of $\nu'_q$ if $p=3$) as $t$ decreases.
Fig.~\ref{figda0p95}(b) also shows that the behavior for larger $t$ is approximately a power law with the exponent $\alpha_2$. 
Thus, we have a second manifestation of the $p=2$ to $p=3$ cross-over.

Following the arguments in Sec.~\ref{sub:rgflows}, what matters when relating the exponent $\nu'_q$ to other exponents according to
Eq.~\eqref{nunuprime} is the leading power law in $L$ at fixed $t$ in the regime where Eq.~\eqref{pq2new} holds. According to our findings
above, this dominant power is indeed always $L^3$ (before the eventual $\phi_q \to 1$ saturation) even though an $L^2$ behavior can also be discerned
for smaller systems if the anisotropy is weak (as in Fig.~\ref{fpanp5}). This proposition is consistent with the form of the divergence of the length
scale $L_c$ with decreasing $t$ in Fig.~\ref{figda0p95}(b), provided that no further cross-over occurs there for even smaller values of $t$. Given that we already
observe a cross-over from the behavior corresponding to $p=2$ in Fig.~\ref{figda0p95}(b) for the larger $t$ values, another cross-over from
$p=3$ back to $p=2$ appears unlikely. The dual power law behavior versus $t$ is not accessible for weaker anisotropy, due to limitations on
computational resources, as the cross-over occurs at larger sizes and smaller $t$ with decreasing anisotropy.

Based on all the results presented here, we conclude that systems with any non-zero anisotropy realize the exponent relation Eq.~\eqref{nunuprime}
with the conventional NG exponent $p=2$ changed to $p=3$ by the anisotropy. For weak anisotropy, various cross-over behavior take place that can be
explained by the two competing power laws in Eq.~\eqref{pq2new}, where we expect that the prefactor of the $L^3$ term to increase with the anisotropy
parameter $\lambda$, as will be discussed in more detail in the next section.

\section{Phenomenological justification for anomalous scaling}\label{sec4c}

In past work \cite{LouSan} it was argued that the exponent $\nu_q^{\prime}$ associated with
$p=3$ in Eq.~(\ref{nunuprime}) can be be generated by considering the effects of a finite correlation
length $\xi$ in a large system. In this limit, the system can be viewed as being made up of
cubic ordered blocks of linear size $\xi$. It was argued that the depth of the effective $Z_q$
potential controlling the angular order parameter, which becomes the scaling variable
for the free energy, is controlled by the number $\propto L^3$ of such cubes. This
intuition was developed for the isotropic case  and is not supported by the numerics
giving $p=2$ in this case \cite{Okubo,Huinew} (as also shown explicitly in Fig.~\ref{fpiclock}).
We expect that this description also does not provide the correct picture of anisotropic
systems and instead consider how the arguments discussed in Sec.~\ref{sec:NGcrossover} for
the physics close to the NG point \cite{Chubukov} can be modified in a non-trivial way in
the presence of space-time anisotropy.

Numerous studies of the nature of the defects in the 3D XY model~\cite{XYe1,XYe2,XYe3,Camarda} have shown
that the dominant fluctuations are of the form of vortex loops. As the region of the flow diagram which
we investigate here is primarily of an XY nature close to the NG fixed point (Fig.~\ref{fflows}), it would
be reasonable to expect that these vortex loops would play a major role also in the symmetry
cross-over. A detailed study \cite{XYe2} of the 3D anisotropic XY model shows that the vortex
loops are elongated along the direction of strongest anisotropy, though they renormalize to the
isotropic limit with increasing length scale. However, topological vortex line defects which
span the entire system may still be present. Further support is given to this idea by
experimental \cite{6Clae} and numerical \cite{6Clan} studies of ferroelectrics, which are modeled
by anisotropic $q=6$ clock models. These studies have found that annealing through the critical
point leads to frozen defects which are consistent with vortex lines.

We conjecture that the $p=3$ behavior of the NG to $Z_q$ cross-over is a result of discrete symmetry breaking in the
presence of vortex line defects. While these line defects should not affect the universality class and the conventional
3D XY critical behavior \cite{XYe2}, their combined effects together with the clock perturbation may be
more subtle. It is important to note here that a vortex flux line description would be consistent with our
observation that the exact form of the interaction used in the imaginary time direction does not play
a role in determining the universal $p=3$ behavior, suggesting instead some stable, topological mechanism.

We have incorporated the RG arguments of Ref.~\cite{XYe2} in Appendix~\ref{sec:app} to show that the form
of the transverse susceptibility in Eq.~\eqref{stsus} remains the same also in the anisotropic systems.
This result implies that, to generate the $p=3$ form of Eq.~\eqref{nunuprime} (assuming here that the basic picture
of the role of the NG fixed point remains), we must modify the relationship between the renormalized field
$\bar{h}$ and the bare clock strength $h$.

In the isotropic case, the renormalization of the $\cos(q\theta)$
irrelevant operator under a scale change $a\to \Lambda a$, where $a$ is the lattice spacing, leads to
the field rescaling $h\to \Lambda^{y_q}h$. In the anisotropic case, we can take the renormalization factor to be the geometric mean
of individual rescaling factors $\Lambda_i^{y_{qi}}$, $i=1,2,3$ in the three directions. Since
$\Lambda_1=\Lambda_2 \propto \Lambda_3$, we still can take a common factor $\Lambda$, but consider the possibility
that the effective scaling dimension related to the third direction vanishes. Thus, we conjecture the rescaling
$h\to h(\Lambda^{2y_q}\Lambda^0)^{1/3}=h\Lambda^{2y_q/3}$. If we use this expression in our analysis of Eq.~\eqref{stsus},
we obtain a modified relation between the physical $k$ and $h$, leading to $k\xi\approx\xi^{{y_q}/{3}}$. Using this
equation and expressing $\xi$ and $\xi'_q$ using $\nu$ and $\nu_q^{\prime}$ as before, we recover the equivalence
expressed in Eq.~\eqref{nunuprime} with $p=3$. This construction is admittedly rather contrived and lacks
rigorous justification. Nevertheless, the renormalization $h\to h\Lambda^{2y_q/3}$ appears to be the only way to
obtain $p=3$ without a complete overhaul of the role of the NG fixed point in the cross-over, and line defects
as an underlying cause at least offer a plausible mechanism to investigate further.

Finally in this section we discuss the possible origin of the two different power laws in the scaling
form of $\phi_q$ in the anisotropic systems, Eq.~\eqref{pq2new}. Since both power laws depend on the same scaled temperature
$tL^{1/\nu}$ according to our empirical findings, and the $L^3$ term should have a prefactor dependent on the
anisotropy parameter $\lambda$ (vanishing as $\lambda\to 0$), the simplest scenario is that the original
scaling function in Eq.~(\ref{phiqscale1}) acquires a third relevant argument $\lambda tL^{1/\nu}$, akin to a DI
perturbation but without generating a new length scale;
\be
\phi_q \sim \Phi(tL^{1/\nu},\lambda tL^{1/\nu},htL^{1/\nu^{\prime}_q},hL^{-|y_q|},\lambda L^{-|\mu|}),
\label{phiqscale4}
\ee
where $\mu<0$ is the scaling dimension of the irrelevant anisotropy. Expanding this function to leading order in $h$
as before, the desired result Eq.~(\ref{pq2new}) can be obtained if the first and second argument in Eq.~\eqref{phiqscale4} generate,
respectively, powers $(tL^{1/\nu})^{\nu(2+|y_q|)}$ and $(\lambda tL^{1/\nu})^{\nu(3+|y_q|)}$. This approach also then predicts a specific power
law dependence of the $L^3$ coefficient; $c_3 \propto \lambda^{\nu(3+|y_q|)}$. In practice, this prediction may be difficult to
test, because changing $\lambda$ also leads to a shift in the critical point, and thereby in the non-universal coefficients
of the arguments in the scaling function.

\section{Conclusions}\label{sec5}

We have studied a family of 2D quantum clock models and explored the behavior of the divergent length scale $\xi'_q$ associated
with U(1) symmetry reduction to $Z_q$ in the ordered phase close to the quantum phase transition. We find that, although
the conventional critical behavior remains in the 3D XY universality class, as expected from a mapping between the quantum
model and a classical 3D clock model (which we also demonstrated in detail), the exponent $\nu_q^{\prime}$ differs from
that of the previously studied isotropic classical model. The differences amount to the value of an exponent $p$, which appears
in the relationship between the exponents $\nu$ and $\nu'_q$ governing the two divergent length scales; Eq.~\ref{nunuprime}.
The exponent $p$ is related to the low-energy excitations of the system at the initial stages of the symmetry cross-over,
where the conventional value is $p=2$ but we find $p=3$.

To further investigate the causes of this unexpected difference, we also studied anisotropic 3D classical clock models and showed
that the unconventional behavior is also realized there for any degree of anisotropy. In some cases, we can explicitly observe a $p=2$
to $p=3$ cross-over behavior (while in other cases only the $p=3$ behavior can be accessed in practice), and this cross-over can be
explained by a scaling function. The evolution of the cross-over with the system size and the temperature implies the unconventional value
of $\nu_q^{\prime}$ in the thermodynamic limit for any anisotropy. Given the known irrelevance of anisotropy in the standard RG sense, the
outset expectation would instead have been a cross-over from a form with significant but decaying scaling corrections with increasing
system size. In some cases we also do observe scaling corrections, but only on top of the anomalous leading scaling behavior.

We have also shown that the anomalous behavior is robust to the precise form of the kinetic terms in the quantum clock
model and the interlayer interactions in the classical models. Based on our observations, we have constructed a phenomenological
treatment to generate the observed scaling exponents. Our scenario implies a non-trivial interplay between the irrelevant
anisotropy operator and the dangerously irrelevant operator responsible for the U(1) to $Z_q$ cross-over. Physically, we propose
that this effect may be related to previously observed \cite{XYe2,6Clae,6Clan} topological vortex line defects in the presence of
anisotropy, and it would be important to further develop this picture with effective models and further studies of the quantum
and classical phase transitions.

As discussed in detail in Appendix \ref{sec6},
in the quantum clock models we did not detect any anomalous finite-size scaling of the critical domain wall energy, which had been
previously found in the $S=1/2$ $J$-$Q$ quantum spin models \cite{Hui15,NDQC}. From the perspective of a scaling function with two relevant
arguments, the anomalous domain wall energy follows from a specific limiting behavior when the scaling arguments become large.
Thus, the quantum clock and $J$-$Q$ models differ in the asymptotic behaviors of the scaling function. The anomaly we found here
for the relationship between $\nu$ and $\nu_q'$ is not related to these asymptotics, but to the altered influence of the NG fixed
point on the renormalization of the clock perturbation in the presence of space-time anisotropy.

The length scale $\nu'_4$ has been studied within the quantum $J$-$Q$ models \cite{Lou09}, but the angular order parameter used
in the past study may not be suitable in the regime where the $Z_q$ fixed point is approached, similar to what we have commented here
on some earlier studies of the classical 3D clock models \cite{note}. It is therefore important to repeat the studies of emergent
symmetry in the $J$-$Q$ models, and to also study the scaling dimension $y_3$ (on the honeycomb lattice \cite{Pujari13,Pujari15})
and $y_4$ (square lattice \cite{Sandvik07,Lou09,NDQC,Jiang08}) of the U(1) breaking perturbation at the critical point. The obvious
question here is whether the conventional NG exponent $p=2$ applies, or whether the space-time anisotropy may change the
scaling to some other value of $p$ as in the quantum clock models. We are planning such studies in the near future.

Interesting future directions also include studies of the square-lattice 2D quantum dimer model \cite{Kivelson87}, where
the exactly soluble Rokhsar-Kivelson (RK) point \cite{Rokhsar88} separates a static staggered VBS phase and a
phase that should have columnar or plaquette VBS order. Studies of the latter phase have reached different
conclusions \cite{Sachdev89,Sylj05,Sylj06,Ralko08,Wenzel12,Banerjee16}, which may possibly be a consequence of the
emergent U(1) symmetry expected in this phase close to the RK point. While the quantum dimer model is difficult
to study with QMC simulations, recent progress has been made with an algorithm of the SSE type with ``sweeping
cluster'' updates incorporating the constraints of the dimer space \cite{Yan19a}. It would be interesting to apply this
method to study the emergent U(1) symmetry in the way we have done here. The most recent work on the quantum dimer model
on the square lattice supports a mixed eight-fold degenerate phase \cite{Yan19b} with both columnar and plaquette long-range
order, as had also been observed in a previous study \cite{Ralko08}. It will be interesting to see if a longer length
scale $\xi'$ can be uniquely identified in this case, or if the splitting of the four peaks in the distribution of the angular
order parameter actually corresponds to two cross-over length scales. Similar questions can also be asked within resonating-valence-bond
wave functions, which also have RK-type points \cite{Alet10,Tang11,Patil14} and can be tuned to ordered phase like the
quantum dimer models \cite{Lin12}.

Another promising route to further our understanding of the development and destruction of U(1) symmetry would be
to study the 2D classical clock model in the presence of disorder. The clean 2D classical model hosts a Kosterlitz-Thouless
phase separating the ordered and paramagnetic phases \cite{DIO1,2Dclockan,2Dclocknum,Surungan}. The introduction of disorder is expected to shrink
the intermediate phase and eventually lead to an infinite disorder fixed point separating the ordered and disordered phases \cite{Senthil1D,Carlon},
thus leading to destruction of the U(1) symmetric phase. Some signatures of the infinite disorder fixed point have been observed in
numerical simulations of the disordered XY model~\cite{Hrahsheh}, and it would be interesting to investigate symmetry
cross-overs with the methods applied in this paper.

\begin{acknowledgements}
We would like to thank Deepak Dhar and Rajesh Narayanan for useful discussions.
H.S. was supported by the NSFC under Grants No.~11734002 and No.~11904024.
A.W.S. was supported by the NSF under Grant No.~DMR-1710170 and by Simons Investigator
Award No.~511064. Some of the computational work was performed using the Shared Computing
Cluster administered by Boston University's Research Computing Services. P.P. would like to
thank Institute of Physics, Chinese Academy of Sciences for hosting a fruitful visit and
facilitating collaboration.
\end{acknowledgements}

\appendix

\section{Susceptibility of the \\ classical clock model}\label{sec:app}

Here we provide a short derivation for the expression used for the transverse
susceptibility in Eq.~\eqref{stsus}, following Appendix B of Ref.~\cite{Chubukov} but
providing additional details required to justify an extension of the arguments to the
anisotropic case.

\subsection{Isotropic model}

Assume a 3D XY model at a temperature $T$ slightly below $T_c$. As we want to ultimately study the
clock model, we add a field $h\cos(q\theta)$. To minimize the free energy, we assume that the XY
ordering lies along the angle $\theta=0$ without loss of generality. We use a coarse grained
description of the system in the standard way by defining a continuous complex valued field
$\phi$, with the corresponding action given by
\begin{eqnarray}\label{XYaction}
S &&=\int d^3x \bigl [ \partial^{\mu}\phi\partial_{\mu}\phi^* - m\phi\phi^* + m_4(\phi\phi^*)^2 \nonumber\\
&&~~~- h(\phi^q+\phi^{*q}) - g(\vec{r})\frac{1}{2i}(\phi-\phi^*) \bigr ].
\end{eqnarray}
Here the last term is added in order to reflect a spatially varying transverse magnetic field of
magnitude $g$ and the imposed clock field of strength $h$ is $q$-fold symmetric.

First we consider the above action for $g(\vec{r})=0$ and assume that we are
solving simply for the configuration which minimizes the action and ignore fluctuations.
This minimum would be given by $\phi(\vec{r})=A$, where $A$ depends on $\{m,m_4,h\}$.
Because we will be considering only small transverse fluctuations in $\phi(\vec{r})$,
we can absorb $A$ into the definitions of $\{m,m_4,h\}$ and set the configuration
(to be the reference state from now on) to be $\phi(\vec{r})=1$.

As we want to study the response to an infinitesimal transverse field at some momentum $\mathbf{k}$,
we now consider Eq.~\eqref{XYaction} for $g(\vec{r})=g\cos(kx)$, where $x$ is along an arbitrary
chosen lattice direction as the system is isotropic (and note that now $g$ just denotes the strength
of the perturbation). The dominant response of the
system will be in the same Fourier component, implying that it should be magnitude preserving
to leading order and of the form $\phi(\vec{r})=\exp(ia\cos(kx))$, where $a\ll1$. Using this
expression for the field in Eq.~\eqref{XYaction} gives
\begin{equation}
\begin{split}
S=&\int d^3x [a^2k^2\sin^2(kx)-m-m_4\\
&-h(2-q^2a^2\cos^2(kx))-g\cos(kx)a\cos(kx)],
\end{split}
\end{equation}
where we have expanded the exponentials where necessary and kept only terms up
to order $a^2$. Performing the integral and assuming $\int d^3x \sin^2(kx)=\int d^3x \cos^2(kx)=C$
(i.e., formally using a system of large but finite size), the $a$ dependent part of the action now is
\be\label{Smin}
S=C(a^2k^2+hq^2a^2-ga).
\ee
As we are only interested in the qualitative behavior of the transverse susceptibility in terms of
$h$ and $k$, $h$ can be redefined to absorb $q^2$. The action above can now be minimized by setting $a=0$
(an unphysical solution) or $a=g/(k^2+h)$ (the physical solution). This result implies
$\phi(\vec{r})=1+ig(k^2+h)^{-1}\cos(kx)$ for small $a$.
The linear response at a particular position is 
\be
\chi(\vec{r})=\lim_{g\to 0}\frac{\phi(\vec{r})_{g}-\phi(\vec{r})_0}{ig}.
\ee
The analysis presented above thus implies
\be
\chi(\vec{r}) \propto \frac{1}{k^2+h}\cos(kx).
\ee
Taking the Fourier component of this at wavenumber $k$ leads to the expression
presented in Eq.~\eqref{stsus}. It must be noted here that we have worked with
fluctuations neglected, and the fluctuations can be taken into account by a renormalization
of the variables $k,h$ by replacing the microscopic length scale by the correlation
length \cite{Chubukov}. The properly renormalized variables are denoted by $\bar{k},\bar{h}$
in Eq.~\eqref{stsus}, and their relation to physical variables are discussed in greater
detail below the equation.

\subsection{Anisotropic model}

We can extend the above analysis to the limit of strong anisotropy by incorporating
the renormalization group elements discussed in Ref.~\cite{XYe2}. Following their notation,
we quantify the anisotropy using $\gamma_0^{-2}=J_{\bot}/J_{\parallel}$, where we
are interested in bare couplings with $\gamma_0^{-2}\gg 1$. Eq.~(C18) of Ref.~\cite{XYe2}
shows that $\gamma_0^{-1}$ is irrelevant under RG flow and asymptotically approaches the
fixed isotropic point $\gamma^{-1}=1$. In the analysis following Eq.~(C18), the authors show that
the flowing value of the anisotropy at scale $l$ is $\gamma_l^{-1}=1+ce^{-l}$, where
$e^l=a_l/a_0$, and $a_l$ ($a_0$) corresponds to the renormalized (bare) length scale.
As they also show that the renormalized isotropic interaction is  $J=\sqrt{J_{\bot}J_{\parallel}}$,
the flowing forms of the interaction strengths can be written as
\begin{mathletters}
\begin{eqnarray}
J_{\bot}^l&=&J \left (1+c\frac{a_0}{a_l} \right ), \\
J_{\parallel}^l&=&J \left (1+c\frac{a_0}{a_l} \right)^{-1}.
\end{eqnarray}
\end{mathletters}  

The above analysis shows how the kinetic term in the action,
Eq.~\eqref{XYaction}, has to be modified to accommodate the anisotropy.
We can see that the form of the action Eq.~\eqref{XYaction} will
depend on which direction we choose for the transverse field
variation. We will label the axes as $x,y,z$, where $z$
corresponds to imaginary time and $x,y$ correspond to space. We first
investigate the effect of an applied transverse field
$g(\vec{r})=g\cos(kz)$. We set the scale
of $J=1$, so that the coefficients in front of the kinetic term
$\partial^z\phi\partial_z\phi$ is $(1+c{a_0}/{a_l})$ and for $x$ and $y$
is $1/(1+c{a_0}/{a_l})$. As the transverse field is assumed to be varying
only in $z$, the $x$ and $y$ kinetic terms are irrelevant. Once again, we
assume a response of $\phi$ to be at the same wavelength as $g(\vec{r})$ and
minimize the action as done earlier for the isotropic case. This gives a final
transverse susceptibility 
\be\label{etranperp}
f_{\bot}(k,h)=\frac{1}{(1+c{a_0}/{a_l})k^2+h}.
\ee
If we were to perform the same analysis using a transverse field varying in
the $x$-direction ($x$ and $y$ directions are equivalent), we would get a
similar form for $f_{\parallel}(k,h)$, with the $k$-dependent part modified
such that
\be\label{etranpar}
f_{\parallel}(k,h)=\frac{1}{(1+c{a_0}/{a_l})^{-1}k^2+h}.
\ee
Note that for both of these forms of the transverse susceptibility, the
asymptotic behavior for a crossover between $h$ and $k$ can still be considered
to be asymptotically controlled only by $k^2$ because $a_0/a_l \to 0$. In this way, other
than non-universal behavior generated before the approach to the asymptotic limit,
the transverse susceptibility is identical to that of the isotropic
case.

\section{SSE Cluster algorithm for\\ quantum clock models}
\label{sec:appqmc}

The SSE QMC simulations used in this work are carried out using the standard series expansion 
of the partition function, $Z={\rm Tr}[\exp(-\beta H)]$;
\be
\label{sseexp}
\exp(-\beta H)=\sum_n \frac{\beta^n}{n!}(-H)^n.
\ee
We express this expansion in the classical clock basis with
states denoted $|A_0\rangle$. By expanding $(-H)^n$ into strings of all the combinations of $n$ of the individual one- and two-body
terms $-Q_i$ and $-V_{ij}$ of $H$ defined in Eq.~\eqref{qham}, $Z$ is translated into a summation over diagonal matrix elements
of the type $\bra{A_0}...V_{i,j}Q_kQ_lV_{m,n}Q_p...\ket{A_0}$ of all operator strings. The matrix elements can be easily evaluated, as
can their changes when substituting a group of operators with other operators. MC sampling is carried out in this configuration
space of strings of all lengths $n$ (where in practice the range of $n$ is bounded because $\langle n\rangle = \beta\langle H\rangle$)
to evaluate operator expectation values of interest. The original variant of this technique for Heisenberg and related models is discussed
in detail in Ref.~\cite{Sandvik10}. The cluster update introduced here is a generalization of that developed for the TFIM in Ref.~\cite{SSE}.

We include a shift in the diagonal terms in Eq.~\eqref{qham}, defining $V_{i,j}=s[\cos(\theta_i-\theta_j)+1]$ to ensure that all
matrix elements of these operators are positive or zero. With our definitions of models (1)-(3) of the off-diagonal terms after Eq.~\eqref{qham},
the operators $Q_i$ also have no negative matrix elements. There are then no negative signs in the expansion Eq.~\eqref{sseexp} of $Z$, as the minus
sign in Eq.~\eqref{sseexp} and those in front of the $V$ and $Q$ terms in the Hamiltonian Eq.~\eqref{qham} cancel.

To execute an ergodic MC simulation
which samples all strings with the appropriate probabilities, we must construct configuration updates which efficiently sample the operator strings
under the periodicity condition where the current state $|A_0\rangle$ is propagated by the current operator string into itself without encountering
any operation with vanishing matrix element.
Each allowed configuration is similar to a term in a path integral, which can be seen more explicitly by inserting a complete
set of states between every pair of consecutive operators in the string. By the definition of the operators $V_{ij}$ and $Q_i$,
only a single basis state contributes at each ``slice''. This generates a set of states $|A_{\alpha}\rangle$, $\alpha=1,\ldots,n$, starting
with the rightmost $|A_0\rangle$ and with each consecutive one generated by acting with the operator $O_{\alpha} \in \{V_{ij},Q_i\}$
indexed by its order $\alpha$ in the string; $|A_{\alpha}\rangle=O_\alpha |A_{\alpha-1}\rangle$ in a simplified notation where we do not
include the normalization factors that give the overall weight $W(A_0,S_n)$ of the configuration;
\be \label{wdef}
W(A_0,S_n)=\frac{\beta^n}{n!} \prod_{\alpha=1}^n \braket{A_{\alpha}|O_{\alpha}|A_{\alpha-1}},
\ee
where $S_n$ denotes an operator string $O_n\cdots O_2O_1$ of length $n$. We again note the ``time'' periodicity, which implies that
$W(A_0,S_n)=W(A_\alpha,S_n^\alpha)$, where $S_n^\alpha$ is obtained from $S_n$ by cyclically permuting the string by $\alpha$ operators.

A simple local update which is traditionally used when sampling these operator strings is to choose an element from the
operator set $\{O\}$  and substitute it in place of one of the operators $O_{\alpha}$ in the current string, ensuring that it
is consistent with the states $\ket{A_{\alpha-1}}$ and $\ket{A_{\alpha}}$. When working with a complete (as opposed to over-complete) basis,
as we do here, this substitution is only possible with maintained constraints if the current and new selected operator are both diagonal,
whence $\ket{A_{\alpha-1}}$ and $\ket{A_{\alpha}}$ are identical. Such an update is therefore called a diagonal update, and is carried out
consecutively for $\alpha =1,\ldots,n$ while storing and updating the state $|A_{\alpha-1}\rangle$ at each step of the process. In
practice, the operator space $\{O\}$ is augmented to include also a number $M-n$ unit operators $I$, which allow for a simple truncation
scheme for the operator sequences, with the maximum expansion power $M$ automatically chosen sufficiently high by the program so as to not
cause any detectable truncation errors. Fluctuations of the number of Hamiltonian operators $n$ are in this scheme achieved through the
diagonal updates without explicitly expanding or shrinking the string.

Updates involving off-diagonal operators must involve at least two operators. Local substitutions of a small number of operators are typically
not very efficient in evolving the configurations. Efficient loop \cite{Evertz,Sandvik10} and cluster updates \cite{SSE} generalizing classical
algorithms such as the Swendsen-Wang scheme \cite{SW} have been developed for many models. In the following, we will present a brief self-contained
description of a cluster update based on a simple extension of the same previously developed for SSE simulations of the TFIM \cite{SSE}. We do not
discuss the overall simulation procedures, as they remain the same as in generic SSE simulations \cite{SSE,Sandvik10}.

\begin{figure}[t]
\includegraphics[width=\hsize]{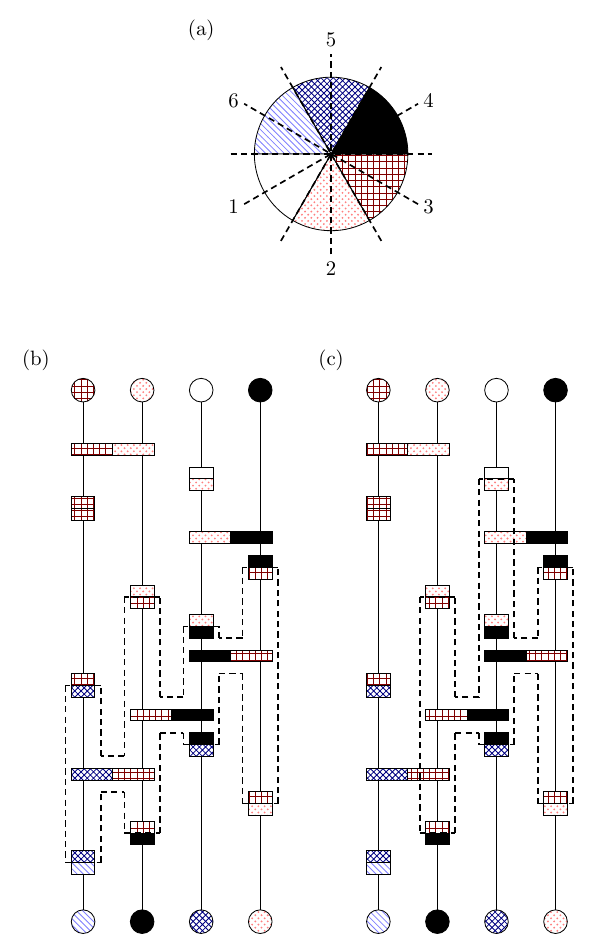}
\caption{(a) The states of the $q=6$ quantum clock model denoted by different
colors and textures, along with all possible axes (lines between the sectors) used
for constructing the Wolff-type clusters. (b) A sample of a section of the operator string
(along with the corresponding boundary clock states represented by circles) with a cluster built
using a simple generalization of the original TFIM algorithm \cite{SSE}. The squares and
rectangles correspond, respectively, to the single-site operators $Q_i$ and the two-body
terms $V_{ij}$. (c) The same section with an instance of a cluster built using the more efficient
method based on a modified Wolff algorithm.}
\label{fcls}
\end{figure}

A direct generalization of the TFIM cluster method suffers from restrictive sampling in the case of the clock model, and we present a modified update
inspired by the Wolff cluster algorithm \cite{Wolff} often used for simulating classical clock models. Pictorial representations of the operator
string and the clusters formed by the two types of cluster updates (simple and Wolff) are shown in Fig.~\ref{fcls}. We specialize to $q=6$ here, and
show in Fig.~\ref{fcls}(a) the color representation we will use for the six possible cyclical states at any lattice point. 
In Figs.~\ref{fcls}(b) and (c), the two types of colored squares and rectangles represent the single-site operator $Q_i$ and the two-site operator
$V_{i,j}$. For ease of visualization, the operator strings shown in Figs.~\ref{fcls}(b) and (c) correspond to a one-dimensional quantum clock model.
The algorithm can be generalized to arbitrary lattices and dimensions with ease.

Figs.~\ref{fcls}(b) and (c) only represent a segment in space and time of a possibly much larger configuration. The colored circles on the top and
bottom signify the boundary states of this segment, i.e., the relevant parts of two propagated states $|A_\alpha\rangle$ and $|A_{\beta}\rangle$.
As the kinetic operators $Q_i$, depicted as squares in Figs.~\ref{fcls}(b) and (c), have off-diagonal elements, different colors are indicated
for the states sandwiching these operator, corresponding to the state before and after the operator has acted. Here it should be noted that, while
the trivial and adjustable diagonal matrix elements of the kinetic operators can be set to zero in principle, in practice we set them to nonzero
values so that the single-site operators can be added and removed in the diagonal updating process. The two-site operators $V_{i,j}$ have only
diagonal elements, and hence always have the same color for the top and bottom part of the operator. These operators instead allow for lateral
difference based on the interaction depending on $(\theta_i-\theta_j)$ (and because of the way we have defined the constant shift of the operator,
the colors on the left and right sides cannot have a difference of $\pi$, i.e. they cannot be on opposite sides of the color wheel shown in
Fig.~\ref{fcls}(a) ).

%These operators instead allow for lateral
%difference based on the interaction depending on $(\theta_i-\theta_j)$ (and because of the way we have defined the constant shift of the operator,
%the colors on the left and right side are always different here though in general this condition is not necessary).

The conventional TFIM update \cite{SSE} generates clusters by considering the squares to act as end-points of a
cluster. The rectangles faithfully extend the reach of the cluster without being modified in the cluster update. Once a
cluster is built using these principles, as shown in Fig.~\ref{fcls}(b), a shift of color can be chosen for the entire cluster,
and picked with appropriate probabilities given by the matrix elements of the $Q_i$ operators involved in the cluster.
This update does not change the relative difference between the colors of the rectangles ($V_{ij}$) involved in the cluster. If the
clusters are large, as is typically the case close to criticality (the region of main interest), the freezing of relative differences
can lead to artificially generated strong correlations over large distances in the sampled states. This limitation does not plague TFIM
simulations, as the restriction to two colors implies that the rectangles only permit a lateral difference of zero.

To allow a more efficient sampling through the cluster update, we incorporate elements of the Wolff cluster update \cite{Wolff} used
in classical spin systems. For a classical clock model, the update is initialized by picking an axis of reflection from the six dashed axes
shown in Fig.~\ref{fcls}(a). This is then set as the axis around which the spins belonging to the cluster are reflected. Starting from a random
spin, the cluster is grown by adding spins to it based on a probability calculated using the energetic gains to the bond connecting the two spin
by reflecting the spin and the temperature at which the system is being simulated. We adapt this to the quantum clock model by choosing an
axis and a starting node which can belong to a $Q$ (square) or $V$ (rectangle) operator. The color of this starting node is reflected about
the axis and the difference generated is recorded. There are two nodes per operator (shown by the two colors), and taking the example of a square
operator and assuming that the starting node is below, the second node is added based on the probability
determined by
\be
\frac{\braket{C^b_r|Q|C^a_r}}{\braket{C^b_r|Q|C^a_o}},
\ee
where b (a) denotes below (above) and r (o) denotes reflected (original). An identical procedure is followed when encountering a rectangular
operator, with the simple modification that above (below) is replaced by left (right). This cluster building technique is now unbiased in its
treatment of $Q$ and $V$ operators, and leads to significant de-correlation in space and imaginary time. The clusters are built probabilistically and are
not completely determined by the operator string structure (the latter which changes only through the diagonal updates). An example of a cluster built
using this technique is shown in Fig~\ref{fcls}(c) using the exact same operator string arrangement as in Fig~\ref{fcls}(b). Note that operators added
to the cluster in Fig~\ref{fcls}(c) are of both $Q$ and $V$ type.

We conclude this Appendix by commenting on a previous generalization of the SSE TFIM cluster update to the quantum Potts model \cite{Ding}.
Rules were implemented for freezing a particular color and permuting the other ones, which is an efficient micro-canonical (weight preserving)
update for Potts models. Such an update would not be weight-preserving for the clock models, where, therefore, the cluster would only be flipped
with some probability that would typically be very small for large clusters.  In contrast, our algorithm for the clock model implements 
the non-preserved weight differently, through the canonical Wolff-like construction described above. The clusters are always flipped with
probability one.

We note that the 3-state antiferromagnetic quantum Potts
model studied in Ref. \cite{Ding} maps onto a 6-state quantum clock
model (while the ferromagnetic system maps onto the 3-state
clock model), and it would be interesting to compare the
efficiency of the clock and Potts cluster algorithms in
this case.

\section{Conventional 3D XY criticality}
\label{sec:app2}

\begin{figure}[t]
\includegraphics[width=\hsize]{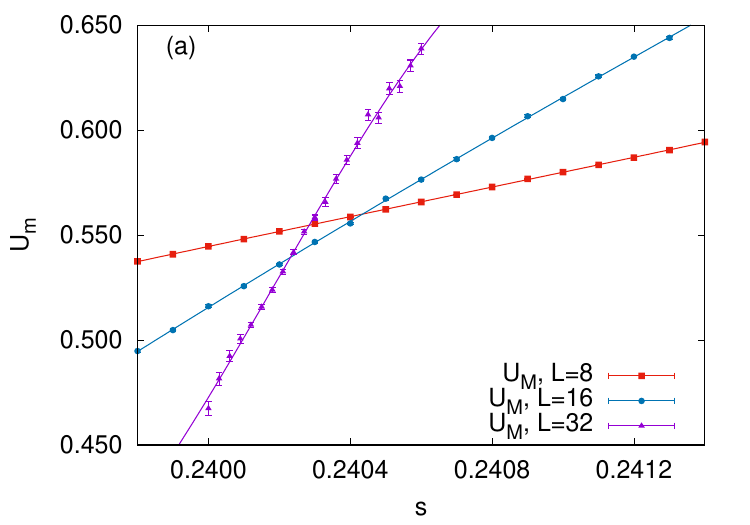}
\includegraphics[width=\hsize]{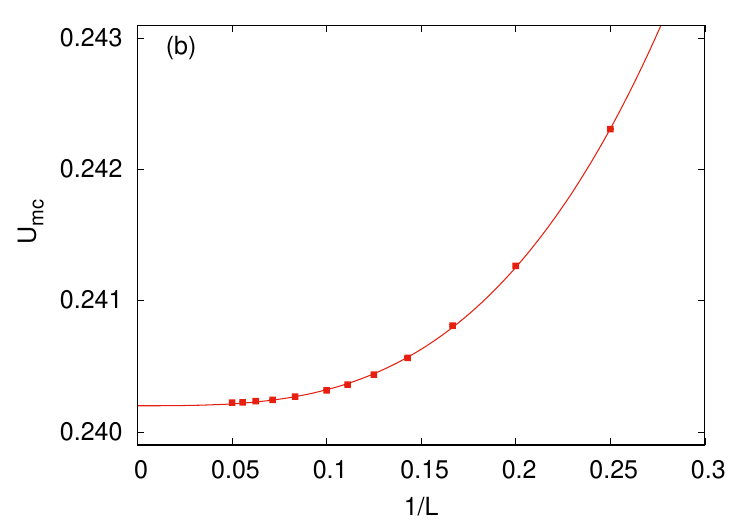}
\caption{(a) Binder cumulant vs the control parameter $s$ for three sample sizes of the quantum clock model
of type (3), which has Potts-like kinetic terms. (b) Cumulant crossing points $U_{mc}$ defined with system sizes $L$ and 
$2L$, extrapolated to the thermodynamic limit value $s_c$ using the expected form $U_{mc}(L)=s_c+a/L^b$. The result is
$s_c=0.240199(5)$.}
\label{fcross}
\end{figure}

\begin{table}
 \begin{tabular}{||l l|l l||}
 \hline
 Quantum& ~~~~~$s_c$ & Classical& ~~~~~$T_c$ \\ [0.5ex]
 \hline\hline
 Model (1)~~~      & ~~0.340450(7) & $\lambda$=0.00    & ~~2.20201(1) \\
 \hline
 Model (2)~~~     & ~~0.50149(6)  & $\lambda$=0.50   & ~~1.697(1) \\
 \hline 
 Model (3) $q=6$      & ~~0.240199(5) & $\lambda$=0.90   & ~~0.8945(2) \\
 \hline 
 Model (3) $q=5$      & ~~0.23909(9) & $\lambda$=0.95   & ~~0.683(2) \\
 \hline
	       &             & $\lambda$=0.90, Potts   & ~~0.75(1) \\
 \hline
 \end{tabular}
 \caption{Critical points for all the classical and quantum models studied in this paper, extracted
 using the technique illustrated in Fig.~\ref{fcross}. We use $q=6$ unless otherwise mentioned. The
 Potts cases signifies an effectively Potts-like interaction in the third direction, while in all other
 classical models a conventional clock interaction was used in all directions.}
\label{tabsc}
\end{table}

\begin{figure}[t]
\includegraphics[width=\hsize]{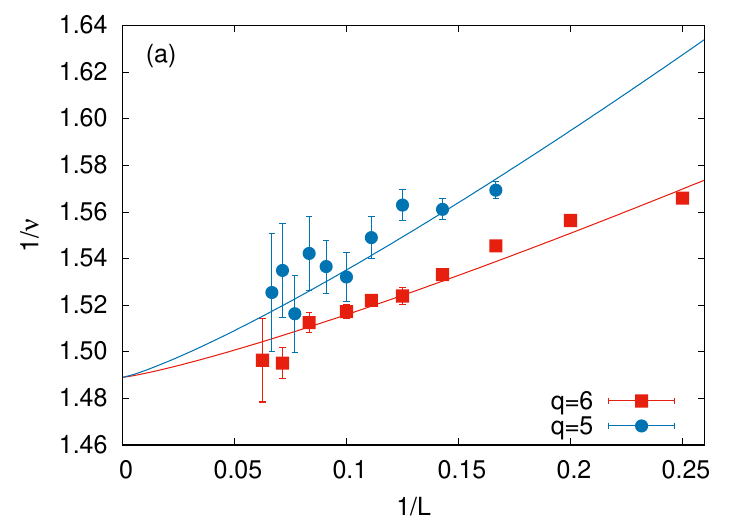}
\includegraphics[width=\hsize]{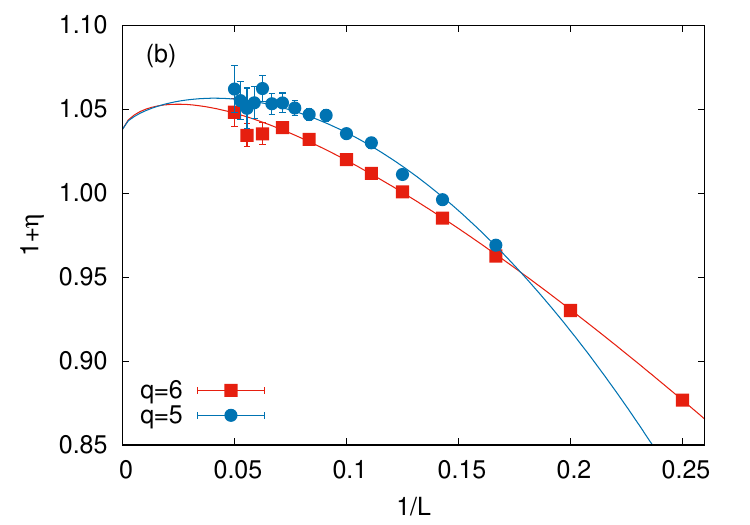}
\caption{(a) Floating exponent $1/\nu$ (defined using the Binder cumulant slopes for system sizes $L,2L$) graphed versus the inverse
system size for quantum clock model (3). The curves show fits in which the known $L \to \infty$ values of the exponents were imposed
and a correction of the expected form $a/L^b$ was used. In (b), similar results are shown for $1+\eta$ (defined using the squared 
magnetization). Here the fitted corrections are sums of two power laws in both cases, to account for the plausible non-monotonic
behavior \cite{Ma18} and the known value of the exponent $\eta \approx 0.038$.}
\label{fc3exp}
\end{figure}

The critical temperatures $T_c$ for the classical models and the tuning parameters $s_c$ for the quantum models were extracted
using size dependent crossing points of the Binder cumulants $U_m$ defined in Eq.~\eqref{umdef} \cite{Luck}. The cumulant exhibits
a transition between $0$ and $1$ at the critical point, with the transition region
shrinking with increasing system size as $L^{-1/\nu}$ and the slope increasing as $L^{1/\nu}$. Crossing points
for sizes $L$ and $2L$ converge to the critical point with a leading correction of the form $L^{-(1/\nu+\omega)}$,
where the correction exponent $\omega$ also depends on the universality class and is often close to unity.
The Supplementary Materials of Ref.~\cite{NDQC} discusses detailed tests of this procedure for other systems.

Examples of cumulants graphed versus the control parameter $s$ in the quantum clock model (3) defined in Sec.~\ref{sec3}
are shown in Fig.~\ref{fcross}(a), and Fig.~\ref{fcross}(b) shows how interpolated $(L,2L)$ crossing points flow toward the
critical point and can be extrapolated by fitting to a single scaling correction. Here we have excluded small system sizes 
until a statistically sound fit is obtained. The critical values $s_c$ extracted using this technique for the three types of quantum
fluctuations which we have used in Sec.~\ref{sec3} are listed Table \ref{tabsc}. The table also lists the critical temperatures
for varying values of the anisotropy $\lambda$ in the classical anisotropic clock models discussed in Sec.~\ref{sec4}.

We have also checked the critical exponents $\nu$ and $\eta$ associated with the quantum phase transition, using quantum clock model (3)
with both $q=5$ and $q=6$. Size dependent versions of the exponents are calculated by taking the ratio of the relevant observables---the
slope of the Binder cumulant for $1/\nu$, and the squared magnetization for $1+\eta$ (the latter specifically for a 2D quantum system
with $z=1$)---computed on lattice sizes $L$ and $2L$. Such ratios directly give the floating critical exponents according to formulas
such as Eq.~(\ref{yqldef}).

Results are shown in Fig.~\ref{fc3exp}. Here we test for consistency with 3D XY universality by fixing the $L \to \infty$
exponents to their known values \cite{3DXY}, $\nu=0.6717(1)$ and $\eta=0.0380(4)$, and only fitting the scaling corrections. As there is
no strong reason to doubt the XY universality class, we have not devoted significant computational resources to these calculations, and
the error bars are therefore quite large and the system sizes are only moderate. In the case of $1/\nu$, we observe monotonic approach to
the asymptotic values and fit with a single power law correction, while in the case of $1+\eta$ we need two scaling corrections with
different signs to account for what appears to be non-monotonic approach to the known exponents, with not yet clearly discernible maximum
values developing in the region corresponding to the largest available system sizes. Non-monotonic scaling behavior has recently been studied
in detail in the context of 2D dimerized Heisenberg models \cite{Ma18}, where some static dimerization patterns had previously been argued
to realize a different universality class but large system sizes revealed the expected O(3) universality with competing scaling corrections
leading to non-monotonic behaviors in some floating exponents, including $1+\eta$. Non-monotonic behavior is clearly necessary in Fig.~\ref{fc3exp}(b)
in order to be able to reach the 3D XY exponent when $L \to \infty$, and the flattening out of the $q=5$ data for the largest systems supports
the existence of such a feature.

\section{Domain walls}\label{sec6}

One of the most direct probes of the two length scales is the domain wall energy. It was predicted \cite{Senthil2,NDQC} in the context
of emergent U(1) symmetry that the domain wall energy density $e$ of a system in the thermodynamic limit depends on both length scales
$\xi$ and $\xi^\prime$ (where $\xi'\gg\xi$). The width of the domain wall is given by the cross-over scale $\xi'$; thus, for a system
with a $2\pi/q$ angular difference across a domain walls, the twist per unit length is $\propto \xi'^{-1}$. The excess ground state energy
density corresponding to such a twist is $e \propto \rho \xi'^{-2}$, where $\rho$ is a stiffness constant with standard Josephson scaling
form $\rho \propto \xi^{-(d+z-2)}$ for a quantum system with dynamic exponent $z$. Multiplying $e$ with this $\rho$ by the width $\xi'$ of
the domain wall, we obtain the energy density per generalized area of a domain wall (i.e., per unit of length in the
case $d=2$ of interest here) in $d$ dimensions;
\be
\Delta E \propto \frac{1}{\xi^{d+z-2}\xi'},
\label{deltae}
\ee
which in our $d=2,z=1$ case is $\Delta E \propto (\xi\xi')^{-1}$.

In systems with a single divergent length $\xi$, finite-size scaling is obtained by expressing a singular
quantity of interest in terms of $\xi$ and then letting $\xi \to L$. With two divergent lengths, as in the case
at hand here, the most natural assumption would be that both $\xi$ and $\xi'$ in Eq.~\eqref{deltae} should be replaced
by $L$. This would mean $\Delta E\sim L^{-2}$ from Eq.~\eqref{deltae} for both the 3D classical clock models ($z=0$) and for 2D quantum
systems with $z=1$. This scaling has been verified for the classical 3D clock model \cite{NDQC}. However, for the 2D
$J$-$Q$ model at its putative deconfined critical point an anomalous scaling was found; $\Delta E \sim L^{-a}$ with
$a<2$ \cite{Hui15,NDQC}. This behavior can be explained if only $\xi'$ in Eq.~\eqref{deltae} is replaced by $L$, while
$\xi \sim \xi'^{\nu/\nu'}$ is replaced by $L^{\nu/\nu'}$; thus $\Delta E \sim L^{1+\nu/\nu'}$. While this conjecture has not been
supported by any concrete underlying theory, the consequences of such replacements in other quantities can consistently explain
anomalous finite-size scaling behaviors of the critical $J$-$Q$ model. These observations motivate us to study also the domain wall
energy of the quantum clock models.

\begin{figure}[t]
\includegraphics[width=\hsize]{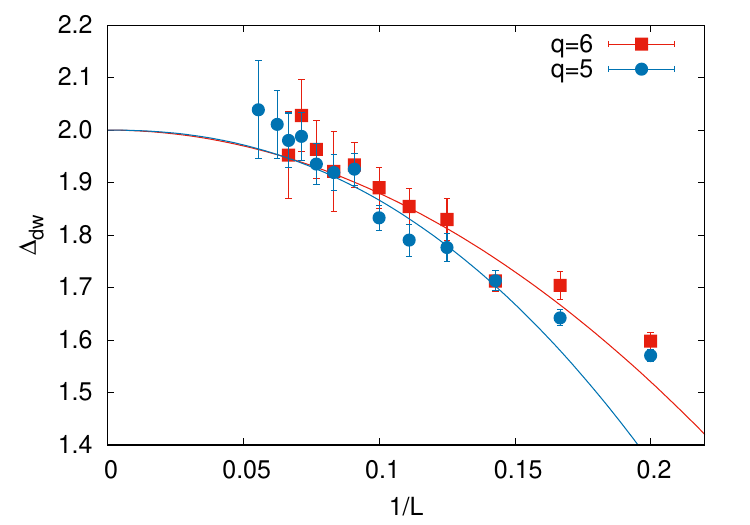}
\caption{Size dependence of the exponent governing the domain wall energy per unit length calculated using the ratio of domain
wall energies for pairs of sizes $(L,2L)$, similar to Eq.~\eqref{yqldef}, with choice (3) of the quantum clock fluctuations.
To test the the finite-size scaling form $\Delta E \sim L^2$, we fix $\Delta_{dw}(\infty)=2$ for $q=5,6$ and fit the finite size
data using a correction of the form $a/L^b$.}
\label{fDWQ6}
\end{figure}

The domain wall energy is calculated by applying fixed boundary conditions in one direction corresponding to one or several
domain walls, $\Delta\theta=n2\pi/q$ with integer $n$, and computing the excess energy over the ground state with
consistent boundary conditions (i.e., $n=0$, with the edges fixed in the same clock direction). In the second lattice
direction periodic boundaries are kept. The corresponding energy density is achieved by normalizing the energy difference by the
length in the periodic (uniform) direction. This quantity $\Delta E$ is determined using QMC simulations carried out at the critical
point and we extract the exponent $\Delta_{\rm dw}$ of the expected power law behavior $\Delta E \sim L^{-\Delta_{\rm dw}}$ 
using finite size scaling.

Ideally, we would like to investigate boundary conditions with a minimum twist $2\pi/q$, but as the domain wall energy is a small
difference between energies which scale with system volume, and has strong relative fluctuations, we use maximum twist conditions.
To this end, for $q=5$ and ($6$) we fixed all $\theta_i=0$ on the left boundary and all $\theta_i=\pi~ (4\pi/5)$ on the right boundary while
maintaining periodic boundary conditions for the top and bottom boundaries, leading to maximum twist. The scaling of the excess energy
is found to be consistent with $\Delta_{dw}=2$, as seen in Fig.~\ref{fDWQ6}. We note that the data for the largest system sizes appear
to fall consistently above the fitted curve, which may possibly indicate some non-monotonicity also in this exponent. However, the error
bars overall are rather large, and the behavior can also be explained by the monotonic form in combination with ``bad luck''.

Thus, for this model, unlike the $J$-$Q$
models, the behavior is consistent with conventional finite-size scaling at a continuous phase transition where both $\xi$ and $\xi^\prime$
are replaced by $L$, leading to an exponent of $2$ from Eq.~\eqref{deltae}. This result supports the notion that the deconfined
quantum critical point of the $J$-$Q$ model realizes a new finite-size behavior (and also finite-temperature properties in the
thermodynamic limit) outside the conventional assumptions of quantum critical scaling \cite{NDQC}.


\begin{thebibliography}{99}
\bibitem{note}
It may appear puzzling why essentially correct results for the exponents $\nu'_q$ were obtained in Ref.~\cite{LouSan} even though
the assumption made for the overall $L$ dependence was incorrect for the regime of $L$ and $t$ used for data collapse as a function 
of $tL^{1/\nu'}$. The reason may be that the exponent $\sigma = \beta/\nu \approx 0.5$ is not very large, and the data collapse was observed 
where $\phi_q$ already approaches its maximal value $\phi_q=1$. In this regime, a different scaling form applies \cite{Okubo,Huinew} (as also 
discussed in the present paper), where the values of $\phi_q$ should not be rescaled at all, i.e., $L^\sigma \phi_q$ with $\sigma=0$ should 
be used. Since the value $\sigma=0.52$ actually used in Ref.~\cite{LouSan} is not very large, the scaling collpase still worked the best when 
$\nu_q$ was close to its correct value. Thus, the reasonable results obtained were in part due to fortuitous circumstances, which may not have 
applied to Ref.~\cite{Zhitomirsky}, where the same assumptions were made.
\end{thebibliography}
\end{document}